# Title: Imaging resonant dissipation from individual atomic defects in graphene

Authors: Dorri Halbertal[1]*, Moshe Ben Shalom[2]*, Aviram Uri[1], Kousik Bagani[1], Alexander Y. Meltzer[1], Ido Marcus[1], Yuri Myasoedov[1], John Birkbeck[2], Leonid S. Levitov[3], Andre K. Geim[2], Eli Zeldov[1]*

Affiliations:
[1]Department of Condensed Matter Physics, Weizmann Institute of Science, Rehovot 7610001, Israel.
[2]National Graphene Institute, The University of Manchester, Booth St. E, Manchester M13 9PL, UK and the School of Physics and Astronomy, The University of Manchester, Manchester M13 9PL, UK.
[3]Department of Physics, Massachusetts Institute of Technology, Cambridge, Massachusetts 02139, USA.
*Correspondence to: dorri.halbertal@weizmann.ac.il (D.H.); moshe.benshalom@manchester.ac.uk (M.B.); eli.zeldov@weizmann.ac.il (E.Z.).

**Abstract**

Conversion of electric current into heat involves microscopic processes that operate on nanometer length-scales and release minute amounts of power. While central to our understanding of the electrical properties of materials, individual mediators of energy dissipation have so far eluded direct observation. Using scanning nano-thermometry with sub-µK sensitivity we visualize and control phonon emission from individual atomic defects in graphene. The inferred electron-phonon "cooling power spectrum" exhibits sharp peaks when the Fermi level comes into resonance with electronic quasi-bound states at such defects, a hitherto uncharted process. Rare in the bulk but abundant at graphene's edges, switchable atomic-scale phonon emitters define the dominant dissipation mechanism. Our work offers new insights for addressing key materials challenges in modern electronics and engineering dissipation at the nanoscale.

One sentence summary: We visualize and control conversion of electric current into local heat at individual atomic defects and identify the dominant dissipation pathway in clean graphene.

Understanding microscopic mechanisms of momentum and energy dissipation is a central problem in the fields ranging from condensed matter to particle physics. It is also of keen interest for designing new approaches to handle, convert, and utilize energy in a bid to address key technological challenges. Dissipation pathways are particularly intriguing in ultra-pure materials of current interest, such as graphene (*1*), because of tight restrictions on the phase space for electron-phonon scattering (*2–4*). Furthermore, unlike particle physics where a particular decay process of interest can be "staged" in a collider, in condensed matter physics we are interested in the processes concealed microscopically within the material. This, along with the minute power released in such processes, poses a key challenge for experimentally probing dissipation at the nanoscale. Here we employ a recently-developed ultrasensitive scanning nanothermometer with sub-µK sensitivity (*5*), achieved with a superconducting quantum interference device placed on an extremely sharp tip—SQUID-on-tip—to probe these subtle effects in high-mobility graphene. Owing to its exceptional cleanness and the two-dimensional nature of its electrons and phonons (*1*), graphene is an excellent platform to study electron-phonon relaxation.

Our measurements were performed on exfoliated graphene encapsulated between hexagonal Boron Nitride (hBN) layers and designed to inject electrons from narrow constrictions into a micron-scale "electron chamber" (Figs. 1A,B and Section 1 (*6*)). Transport measurements in such samples routinely show ballistic signatures over a wide range of temperatures and carrier densities (*7–9*). Our SQUID-on-tip (*10*) acts as an extremely sensitive thermometer (tSOT) (*5*) with an effective diameter of 33 nm and thermal sensitivity of 510 nK/Hz$^{1/2}$ at 4.2 K, and provides an image of the local temperature variations $\delta T(x,y)$ upon scanning at a height $h$ of 10 to 40 nm above the sample surface (as specified). The tip is mounted on a quartz tuning fork (*11*) which allows the tSOT to vibrate parallel to the sample surface with a controlled amplitude $x_{ac} \sim$ 2.7 nm at a frequency of ~37 kHz (Sections 2 and 3 (*6*)). The resulting *ac* signal, $T_{ac}(x,y) = x_{ac} \partial \delta T(x,y)/\partial x$, renders higher sensitivity imaging (Fig. 1C) by avoiding the low-frequency $1/f$ noise of the tSOT. We control the carrier density in graphene globally by a back-gate bias $V_{bg}$ on the Si/SiO$_2$ and locally by applying $V_{tg}$ potential to the tSOT (Fig. 1A). Description of sample fabrication and thermal imaging is provided in Methods (*6*).

Figure 1C shows the thermal signal $T_{ac}(x,y)$ measured while applying a fixed current $I_{dc} = 3$ µA through two of the constrictions as shown in Figure 1B. The image reveals a complex array of fine rings along the edges of the heterostructure (*5*). In addition, three isolated rings are observed in the bulk of graphene labeled 'A', 'B', and 'C'. The bulk rings are rare and have comparable diameters, in sharp contrast to the rings at the edges which are dense and display widely varying sizes (Movies S1-2 (*6*)). We show below that the rings mark dissipation from single atomic defects positioned at their centers.

Electron-phonon cooling pathways in graphene are uniquely interesting for several reasons. Owing to the exceptional stiffness of the carbon bonds, scattering by optical phonons in graphene is inefficient below room temperature. Moreover, the small size of the electron's Fermi surface restricts the phase volume for scattering by acoustic phonons, blocking the intrinsic electron-lattice relaxation pathway (*2–4*). However, theory predict that disorder can significantly ease the electron-phonon scattering (*4*). Our key finding is that hot electrons, generated by the applied current, dissipate their energy through a very specific disorder-induced mechanism: resonant inelastic scattering by local electronic resonances due to individual defects.

Each such localized state (LS) mediates cooling through resonant electron scattering, creating an atomic-scale thermal link between the electronic bath, at an effective hot-electron temperature $T_e$, and the phonon bath, at a base temperature $T_p$ (Section 4 (*6*)). To characterize this novel thermal link we define the "electron-phonon heat conductivity of a defect" $\kappa_{ep}(\varepsilon)$ which describes the power transferred between the baths $P_{ep}(\varepsilon) = \kappa_{ep}(\varepsilon)\Delta T$, where $\Delta T = T_e - T_p$ and $\varepsilon$ is the energy relative to the Dirac point. The resulting *cooling power spectrum $P_{ep}(\varepsilon)$* (CPS)—the fundamental quantity accessed in our experiment through local temperature increase $\delta T(\varepsilon) \propto P_{ep}(\varepsilon)$—is found to peak sharply when the Fermi level $\varepsilon_F$ is aligned with the quasi-bound LS resonant energy $\varepsilon_{LS}$.

Our analysis (Section 4 (*6*)) of the observed resonances suggests that they originate from quasi-bound states arising at a carbon vacancy or adatom bonded to a single C atom when its *sp*² orbital transforms to an *sp*³ state (*12–15*). Such defects are known to produce sharp electronic resonances at energies near the Dirac point (*16*, *17*). While these defects have been extensively investigated by ab initio calculations and STM studies (*12–19*), their prominent role in dissipation has not been anticipated by previous work. The defect-induced CPS originates from the part of local density of electronic states that mediates electron-phonon coupling (EP-LDOS, $D_{ep}(\varepsilon)$) – a quantity that was inaccessible hitherto – convoluted with the electron and phonon Fermi and Bose energy distribution functions (Section 4 (*6*)). Our measurement results and analysis indicate that the spectral width of $D_{ep}(\varepsilon)$ is much greater than the thermal broadening, in which case the CPS can be approximated as $P_{ep}(\varepsilon) \propto D_{ep}(\varepsilon)$ (Section 4 (*6*)). This quantity can be probed experimentally as illustrated schematically in Figs. 2A-C. By parking the tSOT above the defect and varying $V_{tg}$ we can induce local band bending that shifts $D_{ep}(\varepsilon)$ with respect to $\varepsilon_F$ (Section 6 (*6*)). The resulting variation in the measured temperature $\delta T(V_{tg})$ provides a spectroscopic measurement of $P_{ep}(\varepsilon) \propto D_{ep}(\varepsilon)$ (Fig. 2B). Additional information can be obtained by tuning $\varepsilon_F$ by the back-gate $V_{bg}$, yielding resonant peaks in $D_{ep}(\varepsilon)$ aligned as diagonal lines in the $\delta T(V_{tg}, V_{bg})$ map as pictured in Fig. 2C. Importantly, in this configuration phonon emission can be turned on and off by applying a potential on the tSOT tip, representing a new instance of controlling non-equilibrium dynamics and probing it at the nanoscale.

The experimental $\delta T(V_{tg}, V_{bg})$ map of defect 'C', presented in Fig. 2D, displays, in agreement with the discussion above, a sharp resonance (Fig. 2E) along a single diagonal line, which passes closely to the origin in the $V_{tg}$-$V_{bg}$ plane. This resonance is due to the presence of a LS with a narrow single peak in $D_{ep}(\varepsilon)$ (Fig. 2F) close to the Dirac energy (Section 8 (*6*)). Such LSs give rise to the sharp thermal rings observed in Fig. 1C as follows. In order for the LS to cause inelastic electron scattering its energy level $E_{LS}$ has to be aligned with the energy of the impinging electrons, $E_e \cong E_F$. For a given tip potential $V_{tg}$ this resonant condition occurs when the tip is located at a distance $R$ from a LS (Fig. 3A, Movie S3 and Section 6 (*6*)). As a result, each LS displays a sharp peak in $\delta T(x, y)$ and $T_{ac}(x, y)$ along a ring of radius $R$ as shown in Fig. 3B, describing the ring formations in Figs. 1C and 3C.

The variable tip and back gate potentials also provide the spectroscopic means to extract the energy level and the CPS of the LS. By repeatedly scanning the tSOT along the line crossing the defect and incrementing $V_{bg}$ a bell-shaped resonance trace is obtained (Fig. 4A). The shape and polarity of the trace confirms the electrostatic picture described above (Figs. 4A-C, and Section 7 and Movies S5-S6 (*6*)), and its asymptotic value $V_{bg}^{LS}$ describes the energy level $E_{LS}$ of the LS. Similar information can be obtained by sweeping $V_{tg}$ at various values of $V_{bg}$ (Section 7 and Movies S4,S7 (*6*)).

Our spectroscopic analysis of bulk LSs leads to the following conclusions: *a)* The LS energy resides slightly below the Dirac point, $\varepsilon_{LS} = E_{LS} - E_D \cong -22$ meV, as derived from the analysis of the resonance lines (Section 8 (*6*)). *b)* The spectral width of the CPS, $P_{ep}(\varepsilon)$, is about 13 meV (Fig. 2F and Section 9 (*6*)). *c)* No additional LS energy levels are observed at least 180 meV above and below $E_{LS}$. If present, additional concentric rings and bell-shaped traces would have formed. These are, however, absent for the entire $V_{bg}$ and $V_{tg}$ applied range of $\pm 10$ V (Section 7 (*6*)). *d)* This level spacing puts an upper bound on the spatial extent of the LS of less than 2 nm based on charging energy calculation (Section 7 (*6*)). *e)* The sharp energy level and nanometer spatial extent of the LS closely resemble the characteristic features of atomic defects in graphene derived by ab-initio calculations (*14*) and observed by STM (*16*, *17*) in contrast to more extended non-resonant "puddles" originating from disordered substrate potential (*20*, *21*). *f)* The resonant character of the defects and the energy level close to Dirac point are consistent with the $sp^3$ band vacancy model. In particular, vacancies and monovalent adatoms are known to form LSs at energies comparable with our measured values (*14*, *16*) (Section 4 (*6*)) and thus are strong candidates for the observed bulk defects. *g)* The bulk defects are extremely rare (we found a total of seven such defects in a number of samples) corresponding to an estimated average areal density of $2 \cdot 10^5$ cm$^{-2}$ or volume concentration of $5 \cdot 10^{-5}$ ppm if originating from the parent graphite. Their spectral properties appear to be the same within our experimental resolution (Section 8 (*6*)) pointing to a common chemical or structural origin.

These conclusions are reinforced by examining the LSs along the edges of graphene which display spectroscopic features very similar to those of the bulk defects, albeit with some notable differences. Figures 4D-E show tSOT line scans along the bottom edge of the sample crossing through numerous LSs, while increasing $V_{bg}$ (Movies S6-7 (*6*)). Each LS is visible as a bell-shaped trace similar to the ones in Figs. 4A-C, indicating the same microscopic origin of different LSs. In contrast to LSs in the bulk, the edge LSs are extremely dense with some adjacent states only about 1 nm apart (Fig. S15E in (*6*)). This puts an additional bound on their spatial extent (Section 10 (*6*)). We ascribe their origin to the carbon dangling bonds near the edge with high affinity to adatoms and molecules, giving rise to resonance states formed by the resulting $sp^3$ vacancies (*12*–*15*). Notably, quite unlike the LSs in the bulk, the edge LSs display vast variations in $E_{LS}$ values, manifested in Figs. 4D-E by the vertical spread of the bell-shaped traces over entire $V_{bg}$ range of 20 V. This translates into 260 meV spread in $E_{LS}$, limited by our bias range. Remarkably, we resolve states that are less than 2 nm apart, but differ in their $E_{LS}$ by as much as 160 meV (Section 10 (*6*)). Such large energy-level variation may arise from a more diverse chemical origin of the atomic defects at graphene edges as compared to the native bulk defects. An alternative explanation is Coulomb interaction between the charged defects. Indeed, charging a LS by one electron would shift the energy of a neighboring LS at a distance of 2 nm by ~240 meV (Section 7 (*6*)) which agrees with our observations.

The above results suggest that hot electrons lose most of their energy to phonons at the edges of graphene. To verify this conjecture we measured the bare $\delta T(x,y)$ in the absence of tSOT electrostatic influence ($V_{tg}$ at a flat-band condition, see Section 5 (*6*)). Measurement on Sample 2 (Fig. 4F), presented in Fig. 4G, reveals enhanced temperature along the sample edges as compared to a relatively lower temperature in the sample bulk. Corroborated by numerical simulations (Section 5 (*6*)), this finding implies that the LSs along the graphene edges are the dominant source of dissipation at all doping levels reachable in our experiment. Therefore, the excess phonons, corresponding to the overall temperature rise, are not generated locally in the graphene bulk. Instead, the phonons are predominantly emitted through inelastic electron scattering by those LSs at graphene edges that are at resonance at a given $V_{bg}$ value (Section 5 (*6*)). The observed atomic-scale resonant LSs thus emerge as the leading mechanism of dissipation in clean graphene, each acting as point-like lighthouse emitting phonons which then propagate ballistically throughout the entire sample.

The observation of sharply localized resonant states and unveiling their role in dissipation should have significant implications for thermal (*22*), magnetic (*12*, *16*, *23*), chemical (*24*) and transport (*25–28*) properties of graphene. These states are distinct from the extended edge states anticipated for crystalline graphene edges (*23*). Further, resonant dissipation is completely different from the conventional non-resonant picture of electron-phonon coupling (*2–4*), posing interesting questions for future experimental and theoretical work. The new dissipation mechanism may affect the edge transport characteristics (*29–31*) and explain previous observations of the mean free path being limited by the device size in state-of-the-art encapsulated graphene (*7–9*). The resonance states, localized at the edge and in the bulk, thus emerge as a key factor governing the dissipation and possibly limiting the carrier mobility in pure graphene.

**Acknowledgements:**

We thank M. E. Huber for SOT readout setup and M. Solodky for data analysis assistance. This work was supported by the Minerva Foundation with funding from the Federal German Ministry of Education and Research, by the NSF/DMR-BSF Binational Science Foundation BSF No. 2015653 and NSF No. 1609519, by Weizmann – UK Making Connections Programme, and by Rosa and Emilio Segré Research Award. A.K.G. and M.B. acknowledge support from EPSRC - EP/N010345/1, and from the European Research Council ARTIMATTER project - ERC-2012-ADG. L.S.L. and E.Z. acknowledge the support of the MISTI MIT-Israel Seed Fund.

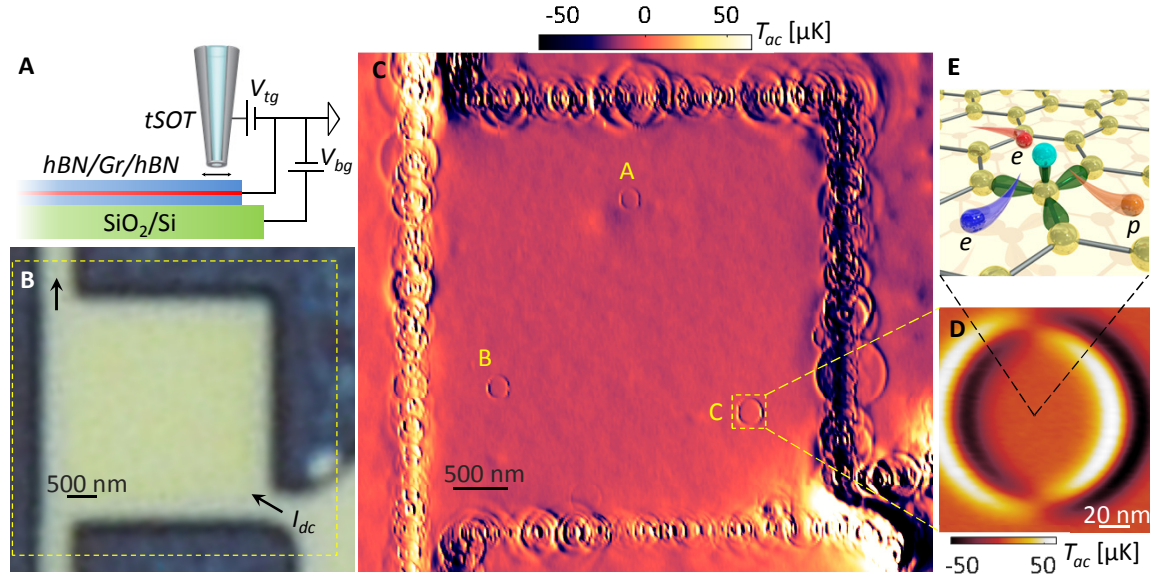

**Figure 1. Observing individual dissipation sources in a graphene heterostructure. (A)** Schematic side view of the measurement setup with hBN-graphene-hBN heterostructure and SQUID-on-tip nanothermometer (tSOT). **(B)** Optical image of the device patterned into 4×4 µm² square chamber (bright). A fixed current $I_{dc} = 3$ µA is driven through the connecting constrictions (arrows). **(C)** Scanning *ac* nano-thermometry $T_{ac}^{TF}(x,y)$ of the area outlined in (B) at $V_{bg} = -2$ V and $V_{tg} = 9$ V at 4.2 K. The tSOT is scanned while oscillating with amplitude $x_{ac} = 2.7$ nm at 12° to $x$ axis. Scan area 5.5 × 5 µm², pixel size 18 nm, scan-speed 20 ms/pixel, $h = 20$ nm, $I_{dc} = 3$ µA. The sharp rings (marked 'A','B','C') uncover three isolated sources of dissipation in the bulk of graphene in addition to a dense array of resonances along the graphene edges. **(D)** Zoom-in on defect 'C' at $V_{tg} = 5$ V. Scan area 140 × 150 nm², pixel size 1.9 nm, scan-speed 20 ms/pixel, $h = 20$ nm, $I_{dc} = 3$ µA. **(E)** Schematic image of an atomic defect in graphene that creates a localized resonant dissipative state at the center of the ring (D) mediating inelastic scattering of impinging electron (red) into a phonon (orange) and a lower energy electron (blue). The defect, forming an sp³ orbital (green), can arise from carbon vacancy, adatom, or admolecule.

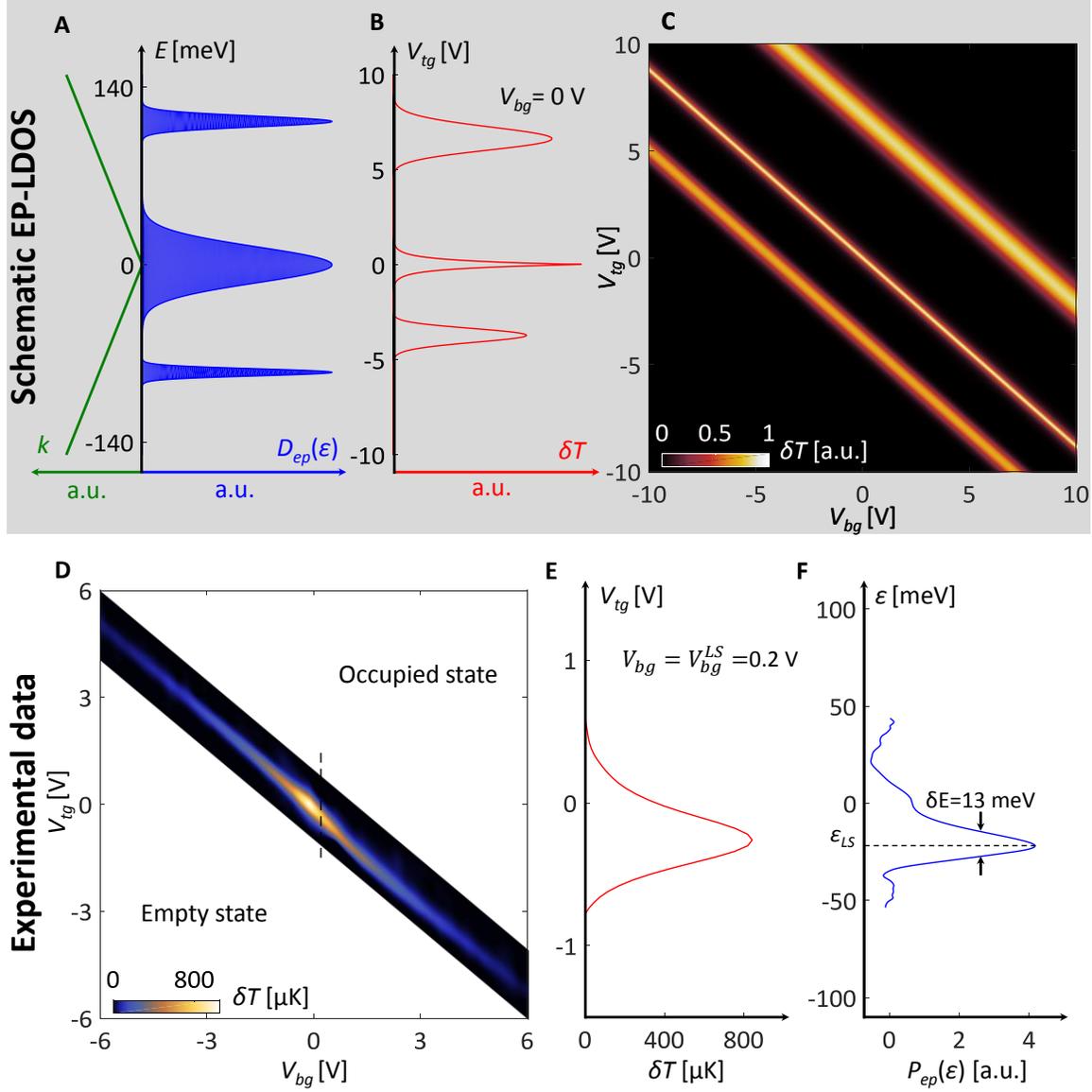

**Figure 2. Thermal nano-spectroscopy of dissipative localized states. (A)** Schematic spectrum of EP-LDOS $D_{ep}(\varepsilon)$ (blue) and the graphene Dirac dispersion relation (green). **(B)** Temperature variation $\delta T(V_{tg})$ measured above the defect vs. tip potential $V_{tg}$ provides nanoscale spectroscopy of the cooling power spectrum $P_{ep}(\varepsilon)$ by tip-induced band bending. **(C)** The expected $\delta T(V_{bg}, V_{tg})$ showing diagonal resonance lines that map the peaks in $D_{ep}(\varepsilon)$. **(D)** The experimental $\delta T(V_{bg}, V_{tg})$ measured by the tSOT at $h = 10$ nm above the center of defect 'C' in presence of $I_{dc} = 3$ µA, revealing a single resonance dissipation line. Signal obtained through integration of $T_{ac}^{tg}$ over $V_{tg}$ (Section 2 (6)). **(E)** A zoomed-in line cut $\delta T(V_{tg})$ along the dashed line in **(D)** at $V_{bg} = V_{bg}^{LS} = 0.2$ V. **(F)** The cooling power spectrum $P_{ep}(\varepsilon) \propto D_{ep}(\varepsilon)$, derived from measurement of the resonant dissipation ring (Section 9 (6)) showing a single sharp peak near the Dirac point at $\varepsilon_{LS} = -22$ meV.

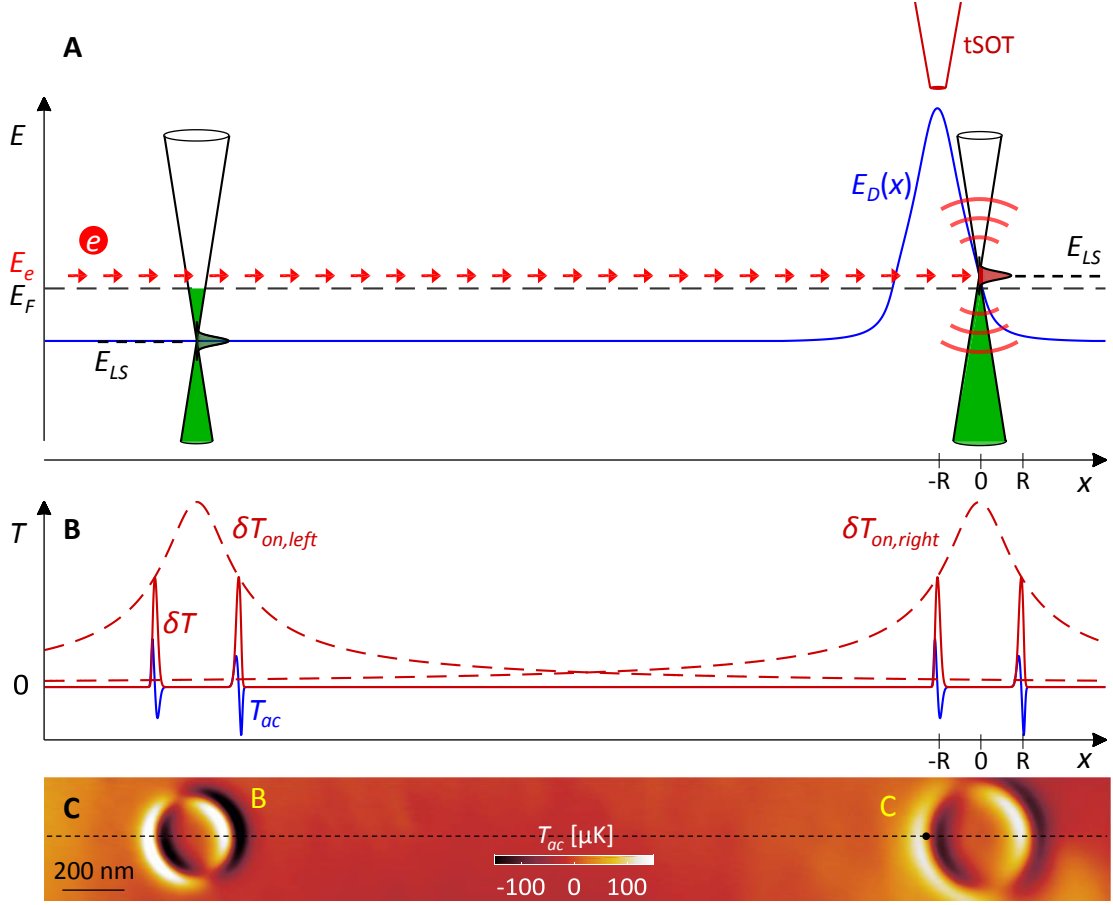

**Figure 3. Origin of the resonant ring structures. (A)** Schematic description of the systems' energy levels corresponding to tip position marked by black point in (**C**): Two localized states with a peak in EP-LDOS at $E_{LS}$ pinned to Dirac energy $E_D$. The tip potential $V_{tg}$ induces band-bending resulting in the calculated position-dependent $E_D(x)$ (blue). The left LS is off resonance while the tSOT positioned at $x = -R$ brings the right LS into resonance for inelastic scattering of electrons injected at energy $E_e$ (red arrows). The result is a point-source of phonon emission (red arcs). The Fermi energy $E_F$ (dashed line) is determined by the back gate voltage $V_{bg}$. **(B)** Dashed line: Schematic temperature profiles due to ballistic 2D phonon emission from LS at resonant conditions, $\delta T_{on}(x) \propto (1 + (x/\ell)^2)^{-1/2}$ with $\ell = 100$ nm (Section 5.2 (6)). Red line: Calculated temperature variation $\delta T(x)$ measured by the scanning tSOT that brings the LS into resonance at a distance $\pm R$ from each defect. Blue: Calculated $T_{ac}(x) = x_{ac} dT(x)/dx$ measured by the tSOT vibrating parallel to the surface. **(C)** $T_{ac}(x, y)$ measured in the central region of Fig. 1C showing dissipation rings around defects 'B' and 'C'. The dashed line describes the scan direction depicted in (**A**). Scan area $3.5 \times 0.4$ µm², pixel size 8 nm, scan-speed 20 ms/pixel, $h = 20$ nm, $V_{bg} = -0.3$ V, $V_{tg} = 5.5$ V, $I_{dc} = 6$ µA and $x_{ac} = 2.7$ nm directed at 18° to the x-axis.

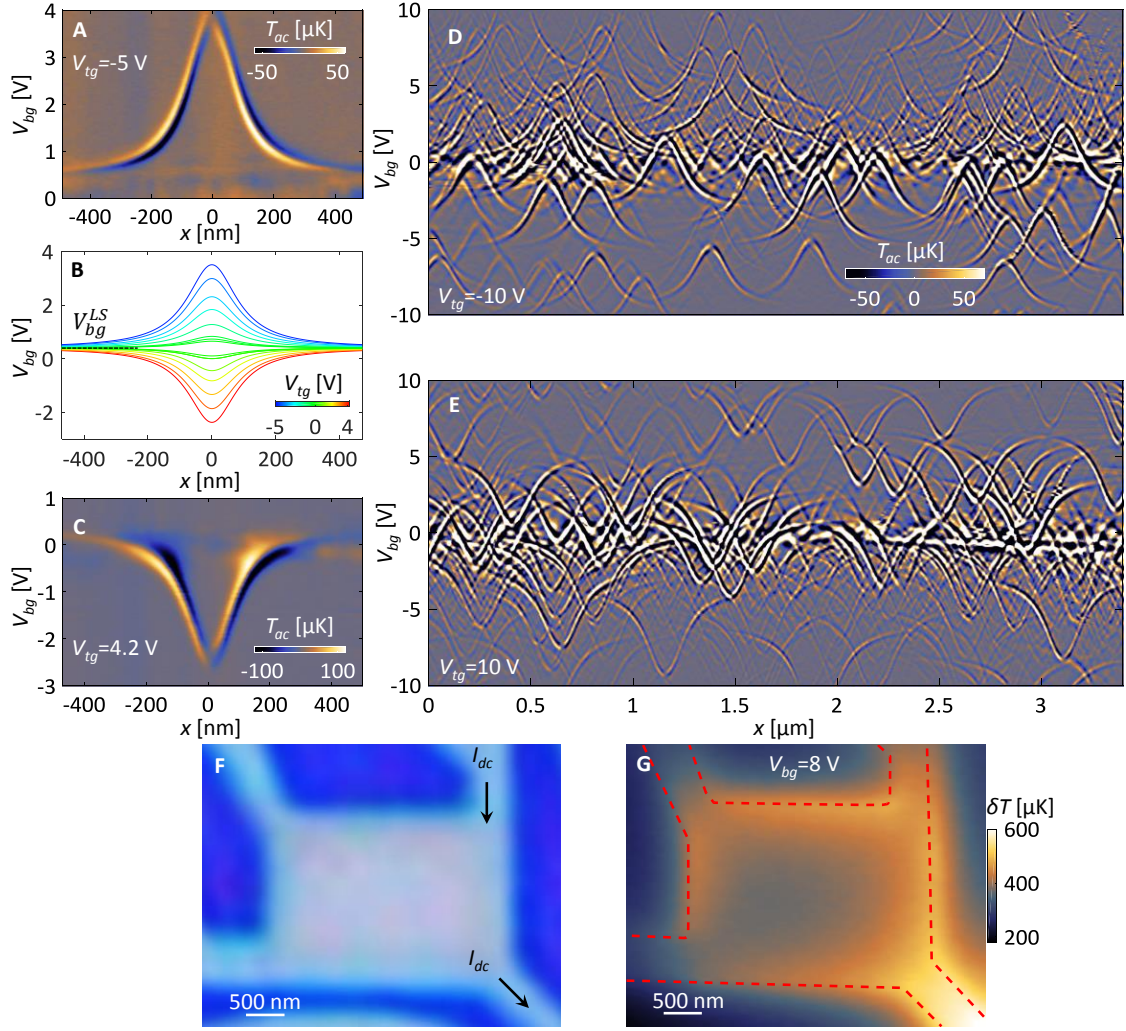

**Figure 4. Thermal spectroscopy of individual bulk and edge localized states.** (**A**) Map of $T_{ac}(x)$ line scans through defect C upon varying $V_{bg}$ at $V_{tg} = -5$ V showing the bell-shaped resonance trace. (**B**) Resonance traces for various values of $V_{tg}$ that switch their polarity at $V_{bg}^{LS}$. (**C**) Resonance trace as in (**A**) at $V_{tg} = 4.2$ V. Scan parameters for (**A-C**): $h = 20$ nm, pixel width 5 nm, pixel height 30 mV, scan-speed 20 ms/pixel, $x_{ac} = 2.7$ nm, $I_{dc} = 3$ µA, linear subtraction was applied to each line to emphasize the resonance traces (see Fig. S11 and Movie S5 for raw images). (**D-E**), Maps of $T_{ac}(x)$ line scans along the bottom edge of the square graphene sample in Fig. 1C at $V_{tg} = -10$ V (**D**) and 10 V (**E**). Each bell-shaped trace originates from a single dissipative atomic defect. Scan parameters: $h = 20$ nm, pixel width 4 nm, pixel height 100 mV, scan-speed 60 ms/pixel, $x_{ac} = 2.7$ nm, $I_{dc} = 3$ µA, high-pass filtering was used to emphasize the pertinent bell-shaped resonance traces. (**F**) Optical image of Sample 2 with patterned hBN/graphene/hBN heterostructure (bright) on SiO$_2$ substrate (dark) and *dc* current $I_{dc} = 1$ µA rms chopped at 35.5 Hz applied as indicated by the arrows. (**G**) Thermal image $\delta T(x, y)$ revealing dissipation along the graphene edges. Scan area 4.7 × 3.7 µm², pixel size 50 nm, scan-speed 200 ms/pixel, $h = 20$ nm, $V_{bg} = 8$ V, $V_{tg} = 0.85$ V corresponding to $V_{tg}^{FB}$ at this $V_{bg}$.

**Supplementary Materials for: Imaging resonant dissipation from individual atomic defects in graphene**

**Materials and Methods**

**1. Sample fabrication and transport characterization**

Our devices are made from single-layer graphene placed atop a relatively thick (~30 nm) crystal of hexagonal boron nitride (hBN), and covered by a thinner (~10 nm) hBN. The heterostructures are assembled using the dry-peel technique on top of an oxidized Si wafer (90 nm of $SiO_2$) which served as a back gate, and then annealed at 300 C in Ar-$H_2$ atmosphere for 3 hours. After this, a PMMA mask is designed by electron beam lithography to define the heterostructures' boundaries and, in a similar step, to define the contacts (~50 nm of superconducting Nb) positions. The exposed heterostructure is etched by a reactive ion plasma combination of $CHF_3$ and $O_2$ which is optimized to uncover a narrow strip (~10 nm wide) of the otherwise encapsulated graphene. This procedure yields low contact resistance (*31*) helping to reduce the local heating at the metal-graphene interface. The graphene/Nb contact interface was designed to be located far outside the imaging chamber (~10 μm away). In the process of defining the imaging chambers geometry the plasma etching conditions were optimized to minimize the uncovered graphene edge to about 1 nm.

The fabrication procedure results in high-quality samples with momentum-relaxation mean free path that is limited by the chamber dimensions over a wide range of carrier concentrations and temperatures (*9*). The studied sample, shown in Fig. 1 and referred to as Sample 1, consisted of a square-shaped chamber of 4 × 4 μm² with three constrictions at the corners connected to external electrical contacts. The three contact points to the chamber were insufficient to perform a four-probe measurement of its transport properties. The top constriction, however, had additional contacts above it, which allowed us to carry out four-probe measurement of its magneto-resistance as a function of $V_{bg}$ as shown in Fig. S1. The quantum Hall oscillation data allow us to extract the back-gate capacitance to the graphene $C_{bg} = 1.25 \cdot 10^{11}$ 1/cm²V and determine the charge neutrality point (CNP) of the constriction, $V_{bg}^{CNP} = -0.26$ V. The CNP of the 500 nm wide constriction, however, may differ from the CNP in the square bulk chamber which we estimate to be $V_{bg}^{CNP} = 0.48$ V as described in Section 8 below. A second sample (Sample 2) of similar structure but with thicker hBN top layer of 30 nm was used for the measurements described in some sections in the SM as noted. Sample 2 had $C_{bg} = 0.9 \cdot 10^{11}$ 1/cm²V and $V_{bg}^{CNP} = 0.6$ V.

**2. tSOT characterization and thermal imaging schemes**

The tSOT was fabricated by self-aligned three-step Pb deposition onto a pulled quartz pipette as described previously (*5*, *10*). Figure S2A shows a SEM image of the tSOT device used in this work for measurements on Sample 1. The I-V characteristics of the tSOT at 4.2 K are shown in Fig. S2C at various applied magnetic fields. The current $I_{tSOT}$ is measured using a SQUID series array amplifier (*11*) (SSAA) as shown in the inset of Fig. S2C. For $I_{tSOT} < I_c(H)$ the device is in the superconducting state, while at currents above the critical current $I_c(H)$ a finite voltage $V_{tSOT}$ is developed redirecting part of the externally applied current to the shunt resistor $R_s$. The interference pattern of the critical current $I_c(H)$ in Fig. S2D shows the central lobe of period 2.4 T corresponding to effective tSOT diameter $d = 33$ nm. The higher lobes are suppressed and truncated by the upper critical field $H_{c2} = 1.73$ T of the Pb film. The magnetic and thermal sensitivities of the tSOT were characterized as described in Refs. (*5*, *10*). The flux noise of the tSOT in the white noise region above about 1 kHz at 4.2 K was 145 n$\Phi_0$/Hz$^{1/2}$ at applied field $\mu_0 H = 0.56$ T, corresponding to spin noise of 0.85 $\mu_B$/Hz$^{1/2}$; the thermal noise at zero field was 510 nK/Hz$^{1/2}$. The thermal coupling between the tSOT and the sample was provided by 60 mBar of He exchange gas (*5*). Sample 2 was studied using a similar tSOT of 48 nm effective diameter.

The tSOT was attached to a quartz tuning fork (TF) as shown in Fig. S2B. The TF was excited by applying an *ac* excitation voltage $V_{TF}$ to it at the resonance frequency $f_{TF} = 36.9867$ kHz and the current through the TF was monitored. As the tip approaches the sample surface to within a few nm, a shift in $f_{TF}$ and in the phase of the current is detected and used as a feedback for height control. After approaching the surface, the tSOT was retracted to a height $h$ of 10 to 40 nm (as indicated in figure captions) above the surface of the hBN, the feedback was switched off, and the scanning was performed at a fixed height $h$.

In order to avoid the $1/f$ noise of the tSOT, in addition to the *dc* $\delta T$ measurement, in our studies we also employed $T_{ac}$ measurements using two different methods. The first method uses the *ac* oscillation of the tSOT, $x_{ac}$, parallel to the sample surface by applying excitation $V_{TF}$ to the TF at its resonance frequency $f_{TF}$. The resulting measured $T_{ac}^{TF}$ reflects the local temperature gradient along the vibration direction of the TF, $T_{ac}^{TF}(x,y) = x_{ac} d\delta T(x,y)/dx$. Figures 1, 3 and 4 show such $T_{ac}^{TF}$ data which we refer to as $T_{ac}$ in the main text for brevity. If $x_{ac}$ is sufficiently small and $T_{ac}^{TF}(x,y)$ is imaged over a sufficiently large area, the *dc* temperature variation can be reconstructed by integration $\delta T(x,y) = \int^x dx' T_{ac}^{TF}(x',y)/x_{ac}$. Such integration, however, is usually unnecessary since the essential information can be readily extracted directly from $T_{ac}^{TF}$. In the second method we apply an *ac* potential $V_{tg}^{ac}$ to the tip in addition to the *dc* $V_{tg}$, which measures the change in the local temperature in response to a change in the tip potential, $T_{ac}^{tg}(x,y) = V_{tg}^{ac} d\delta T(x,y)/dV_{tg}$. By measuring $T_{ac}^{tg}$ vs. $V_{tg}$ at a fixed tip position the *dc* temperature variation $\delta T(V_{tg})$ can be obtained by integration, $\delta T(V_{tg}) = \int^{V_{tg}} dV_{tg}' T_{ac}^{tg}(V_{tg}')$ which provides direct spectroscopic information on the electron-phonon cooling power spectrum (see Sections 4 and 9 below). The data presented in Figs. 2D and 2E were obtained by this method.

Usage of these three measurement methods is demonstrated in Figs. S3B-D, which details imaging of the dissipation ring of a defect on the left edge of Sample 2 (Figs. S3A and S4A). Imaging was done by applying $V_{tg}^{ac} = 0.1$ V rms at $f = 3.06$ kHz and $x_{ac} = 1.5$ nm rms at $f_{TF}$ in the presence of $I_{dc} = 1$ µA rms chopped at 35.5 Hz (see Section 5 below). The $\delta T(x,y)$ in Fig. S3B shows a ring of enhanced temperature, described schematically by the solid red line in Fig. 3B in the main text. Note that the intensity of the ring is higher within the graphene heterostructure region (right) than outside of it (left) due to better heat conductivity of the hBN/graphene/hBN as compared to SiO$_2$ substrate. Figure S3C shows the simultaneously measured $T_{ac}^{TF}(x,y)$, described schematically by the blue line in Fig. 3B. Figure S3D shows thermal imaging using the *ac* tip gating method. Since varying tip potential $V_{tg}$ changes the radius $R$ of the ring, the resulting signal can be expressed as $T_{ac}^{tg}(x,y) = V_{tg}^{ac} dT(x,y)/dV_{tg} = (dR/dV_{tg})V_{tg}^{ac} dT(r,\theta)/dr$, where $r$ is the radial distance from the center of the ring. The two *ac* methods are thus equivalent to measuring two spatial derivatives of the *dc* image, $T_{ac}^{TF}(x,y) \propto dT(x,y)/dx$ and $T_{ac}^{tg}(x,y) \propto dT(r,\theta)/dr$.

In the studies described in SM we have mainly employed the TF *ac* method (Fig. S3B) due to its superior S/N performance and operational convenience. Figure 1C presents such *ac* thermal image $T_{ac}^{TF}(x,y)$ of the sample. Note that the bulk of the sample has a uniform background $\delta T(x,y)$ temperature (which is not affected by the local tip gating) that is slightly higher than that of the surrounding SiO$_2$ substrate (see Section 5 below). This small step-wise change in $\delta T(x,y)$ at the sample edges translates accordingly into a strip of a bright background $T_{ac}^{TF}(x,y)$ signal along the left edge of the sample and a similar negative strip along the right edge in Fig. 1C.

## 3. Calibration of the vibration amplitude of the tuning fork

The vibration amplitude of the tSOT attached to the TF excited at its resonance frequency can be derived by a direct comparison of the $\delta T(x,y)$ and $T_{ac}^{TF}(x,y)$ images. We demonstrate here the procedure by imaging a constriction in Sample 2 as marked in Fig. S4A. A *dc* current $I_{dc} = 4$ µA rms, chopped by a square wave at 35.5 Hz was applied to the constriction as indicated by the arrows. The resulting images of $\delta T(x,y)$ and $T_{ac}^{TF}(x,y)$, measured simultaneously at a flat-band condition $V_{tg}^{FB}$, are presented in Figs. S4B,F.

The chopping of the current allows measuring $\delta T(x,y)$ at a frequency of 35.5 Hz using a lock-in amplifier, thus significantly reducing the $1/f$ noise contribution to the tSOT signal. (The presented $\delta T(x,y)$ has been corrected for the numerical factor of $2\sqrt{2}/\pi$ between the first harmonic lock-in reading of a square wave and the corresponding time averaged value.) The measured $T_{ac}^{TF}(x,y)$ is expected to follow $T_{ac}^{TF}(x,y) = x_{ac}\,\partial\delta T/\partial x + y_{ac}\,\partial\delta T/\partial y = \boldsymbol{r}_{ac}\cdot\nabla\delta T$ where $x_{ac}$ and $y_{ac}$ are the rms values of the tSOT vibration along the imaging axes and $\boldsymbol{r}_{ac}$ is the corresponding vibration vector. In order to derive $x_{ac}$ and $y_{ac}$ we take numerical derivatives $\partial/\partial x$ and $\partial/\partial y$ of Fig. S4B (Figs. S4C,D) and perform numerical two-parameter fit of a linear combination of Figs. S4C,D to the measured $T_{ac}^{TF}(x,y)$ image of Fig. S4F. The resulting best fit numerical image, which is presented in Fig. S4E, provides values $x_{ac} = 4.05$ nm and $y_{ac} = 0.31$ nm, indicating that the TF oscillated at an angle of 4.3 deg with respect to the $x$ axis. This procedure was then repeated for various values of TF excitation voltage $V_{TF}$. Figure S4G shows the resulting linear dependence of the tSOT vibration amplitude $r_{ac} = \sqrt{x_{ac}^2 + y_{ac}^2}$ on $V_{TF}$.

**Supplementary Text**

**4. Theoretical analysis of the electron-phonon cooling power spectrum**

Graphene is known to host a wide variety of atomic-scale defects which can act as resonant scatterers, trapping electrons in quasibound states. Here we show that such defects enhance the inelastic electron-phonon scattering and provide a pathway for electron-lattice energy dissipation. We define the notion of the *electron-phonon cooling power spectrum*, which is the quantity measured by our instrument, and show that in realistic cases it factors into a product of a contribution that depends on the defect density of states and a contribution that depends on the electron and phonon energy distributions. While here, for simplicity, we focus on the two-temperature case ($T_e$ and $T_p$ for the electron and phonon bath respectively), the notion of the cooling power spectrum can be easily generalized to a more general non-equilibrium transport regime. We stress that such resonant cooling pathways have not been studied previously, neither theoretically nor experimentally.

We start with a brief summary of the studies of resonant scatterers in graphene and then proceed to describe inelastic scattering mediated by these defects. Ab initio (*14*) and STM (*16*, *17*) studies have shown that quasi-bound states with energies near the Dirac point arise in a robust manner when adatoms or polar groups like H, F, $CH_3$ or $OH^-$ bind covalently to carbon atoms (Fig. 1E), transforming the trigonal $sp^2$ orbital to the tetrahedral $sp^3$ orbital. Each transformed C atom represents a vacancy in the $\pi$-band that produces a quasibound state i.e. a sharp resonance positioned near the Dirac point. In contrast, the defects having other symmetries (e.g. adatoms positioned between two C atoms or at a hexagon center) typically form resonances far away from the Dirac point. This behavior can be captured using a tight-binding model on the honeycomb lattice with an on-site disorder

$$H_{el} = \sum_{|r-r'|=1} t_0\left(\psi_r^\dagger \psi_{r'} + h.c.\right) + \sum_r u(r)\,\psi_r^\dagger \psi_r, \qquad u(r) = \sum_i u_i\,\delta(r - r_i), \quad \text{(S1)}$$

where $r_i$ are the defect positions, and the energy parameters $u_i$ are on the order or greater than graphene bandwidth, W=$6t_0$ (*15*). Focusing on a single defect placed at the origin $r=0$ on sublattice *B*, and passing to the momentum representation, we obtain

$$H_{el} = \sum_{\boldsymbol{k}} \begin{bmatrix} \psi_{A,k}^\dagger \\ \psi_{B,k}^\dagger \end{bmatrix} \begin{bmatrix} 0 & t_0 f(\boldsymbol{k}) \\ t_0 f^*(\boldsymbol{k}) & 0 \end{bmatrix} \begin{bmatrix} \psi_{A,k} \\ \psi_{B,k} \end{bmatrix} + \sum_{\boldsymbol{k},\boldsymbol{k}'} u_0\,\psi_{B,\boldsymbol{k}'}^\dagger \psi_{B,\boldsymbol{k}}, \quad \text{(S2)}$$

where $f(\boldsymbol{k}) = e^{i\boldsymbol{n}_1 \boldsymbol{k}} + e^{i\boldsymbol{n}_2 \boldsymbol{k}} + e^{i\boldsymbol{n}_3 \boldsymbol{k}}$ with $\boldsymbol{n}_{1,2,3}$ the vectors connecting a C atom to its nearest neighbors, and the sum is taken over $\boldsymbol{k}$ and $\boldsymbol{k}'$ in the Brillouin zone. At low defect concentration and for $u_0$ values large compared to the bandwidth $W = 6t_0$, each defect hosts a single resonance state with the energy $\varepsilon_{LS}$ close to the Dirac point, broadened due to hybridization with the states in the Dirac continuum. Scattering of band electrons on the defect is described by a T-matrix, i.e. the defect potential renormalized by multiple scattering processes:

$$t(\varepsilon) = \frac{\pi v^2}{\varepsilon \ln(iW/\varepsilon) + \delta}, \qquad \delta = \pi v^2/u_0 \ll W. \tag{S3}$$

that features a resonance centered at an energy $\varepsilon_{LS}$, near the Dirac point ($\varepsilon_{LS} \approx -\delta/\ln(W/\delta)$). Here the parameter $\delta$ describes detuning of the resonance from the Dirac point, $W$ is the bandwidth, $W \approx 6$ eV, $v \approx 10^6$ m/s is the Fermi velocity, and for simplicity we set Planck's constant equal unity. At $u_0$ large compared to $W$ the resonance energy $\varepsilon_{LS}$ is nonzero and small compared to $W$. The sign of $\varepsilon_{LS}$ is opposite to that of $u_0$, i.e. for $u_0$ positive $\varepsilon_{LS}$ is of a negative sign. This is in good agreement with the results of ab initio studies (*14*) predicting the resonance energy values: $\varepsilon_{LS} = -0.03, -0.11, -0.70, -0.67$ eV for the $sp^3$ orbital state induced by H, CH$_3$, OH, and F, respectively.

The contribution of a defect to the single-particle density of states is given by

$$\frac{1}{\pi} \operatorname{Im} t(\varepsilon) = \frac{\pi v^2 |\varepsilon|/2}{(\varepsilon \ln(W/|\varepsilon|) + \delta)^2 + (\pi \varepsilon/2)^2}, \tag{S4}$$

where we used an identity $\ln\left(\frac{iW}{\varepsilon}\right) = \ln\left(\frac{W}{|\varepsilon|}\right) + i\frac{\pi}{2}\operatorname{sgn}\varepsilon$. The energy dependence $\frac{1}{\pi}\operatorname{Im} t(\varepsilon)$ defines a sharp peak positioned near the Dirac point. The half-width of the resonance in $t(\varepsilon)$ predicted by the above model is $\gamma \approx \pi\delta/(2\ln(W/\delta))$, i.e. is small when $\delta$ is small.

The spatial behavior of the states associated with such resonances can be understood in a simple way using the low-energy Dirac Hamiltonian

$$H = \begin{bmatrix} 0 & -iv\partial_+ \\ -iv\partial_- & u_0\delta(x) \end{bmatrix}, \qquad \partial_\pm = \partial_x \pm i\partial_y. \tag{S5}$$

For a large on-site energy $u_0$, the Hamiltonian hosts a single zero-energy eigenstate of the form $\psi(x, y) = \frac{a}{x - iy}\begin{pmatrix}1\\0\end{pmatrix}$, wherein the approximate particle-hole symmetry becomes exact in the large-$u_0$ limit. The wavefunction $\psi(x, y)$ features a power-law tail and a log-diverging normalization, and is therefore marginally delocalized. When the defect resides on the graphene *B* sublattice, the zero-mode wavefunction takes nonvanishing values on the *A* sublattice, and vice versa. Interestingly, the zero modes survive when many defects are placed on the same sublattice, *A* or *B* (*15*).

Next we proceed to calculate the cooling power spectrum for a defect and show that in realistic cases it is proportional to the defect density of states with a prefactor that depends on the electron and phonon energy distributions. The contribution of a resonant scatterer to electron-phonon scattering can be understood pictorially as resonant trapping of band electrons on the quasi-bound state with the released energy difference transferred to phonons. Our treatment of defect-assisted electron-phonon scattering accounts for the key aspects of interaction between electrons and phonons in graphene. First, the unusually high energy of optical phonons makes optical phonons irrelevant for electron-lattice cooling at temperatures below few hundred kelvin. As a result, the long-wavelength acoustic phonons play a dominant role despite their relatively weak coupling to electrons. Second, momentum conservation strongly limits the phase volume for electron-phonon scattering in graphene bulk. Atomic defects can absorb phonon recoil momentum and thus relieve the bottleneck due to the small phase volume, giving rise to an enhanced cooling rate near the defect. Microscopically this can be described with the help of the Hamiltonian

$$H = H_{el} + H_{ph} + H_{int}, \tag{S6}$$

where $H_{el}$, given in Eq. (S1), describes electrons and their scattering by the defects as discussed above, $H_{ph} = \sum_k \omega_k b_k^\dagger b_k$ describes acoustic phonons, and $H_{int} = \sum_{p'=p+k} g\sqrt{\omega_k} a_{p'}^\dagger a_p b_k$ + h.c. describes electron-phonon interaction. Here $g\sqrt{\omega_k}$ is the graphene deformation potential matrix element (see Ref. (*4*)). Starting with the Hamiltonian (S6) we calculate the energy dissipation rate as

$$P_{ep}(\varepsilon) = \sum_{p'=p+k} \omega W_{p\to p'k}(1-n')n(N_k+1)\delta(\varepsilon'+\omega-\varepsilon)$$

$$- \sum_{p'+k=p} \omega W_{p'k\to p}(1-n)n'N_k \delta(\varepsilon'+\omega-\varepsilon) \tag{S7}$$

Here $n$, $n'$ and $N_k$ are Fermi and Bose distributions for electrons and phonons with momenta $p$, $p'$ and $k$, and with the energies $\varepsilon = \varepsilon_p$, $\varepsilon' = \varepsilon_{p'}$, $\omega = \omega_k$, respectively. The two terms in (S7) describe processes of phonon emission and phonon absorption. The microscopic transition rate is evaluated from Fermi's Golden Rule as $W_{p \to p'k} = \frac{2\pi}{\hbar}|M|^2$ with the matrix element $M$ describing defect-assisted electron-phonon scattering comprising several different contributions. Namely, phonon emission can take place either before or after the carrier is trapped on the quasibound state (for details see Ref. (*4*)), giving

$$M = M_0(k)\, G(\varepsilon, p' + k)\, \frac{1+\sigma_3}{2}\, t(\varepsilon) + t(\varepsilon')\, \frac{1+\sigma_3}{2}\, G(\varepsilon', p - k) M_0(k)$$
$$+ \sum_q t(\varepsilon')\, \frac{1+\sigma_3}{2}\, G\!\left(\varepsilon', q - \tfrac{k}{2}\right) M_0(k) G\!\left(\varepsilon, q + \tfrac{k}{2}\right)\, \frac{1+\sigma_3}{2}\, t(\varepsilon) \quad \text{(S8)}$$

where $M_0(k) = g\sqrt{\omega_k}$ is the electron-phonon matrix element, $G(\varepsilon, p) = 1/(\varepsilon - v\sigma p)$ is the graphene electron Greens function, with the 2x2 matrices accounting for the A/B sublattice structure. The quantity $t(\varepsilon)$ is the scatterer T-matrix evaluated above, the factors $\frac{1+\sigma_3}{2}$ account for the sublattice A or B on which the defect is located.

Starting from a completely general expression (S8), our subsequent analysis proceeds differently for the cases when the values of temperature $T$ and bias $eV$ are (i) much smaller, or (ii) much greater than the resonance width $\gamma = \pi\delta/\bigl(2\ln(W/\delta)\bigr)$. In the case (i) the 2x2 matrix structure in (S8) simplifies after taking into account that the change of electron energy—which is of the order of $k_B T$ and $eV$, whichever is greater—is small on the scale of $\gamma$ and thus the process is quasi-elastic. Also, at not too low temperatures the phonon momentum values $k$ are typically large compared to electron $p$ and $p'$ values, allowing us to replace $G(\varepsilon, p)$ with $\pm \frac{1}{v\sigma k}$ in the first two terms and giving an expression

$$M(k) = M_0(k)\, \frac{i\sigma \times k}{vk^2}\, t(\varepsilon) + \frac{1}{2\pi v^2}\, M_0(k)\ln\!\left(\frac{W}{vk}\right) \frac{1+\sigma_3}{2}\, t(\varepsilon')t(\varepsilon). \quad \text{(S9)}$$

For temperatures of interest the phonon momentum $k_p \approx k_B T_e/s$, where $s \approx 2 \cdot 10^4$ m/s, the sound velocity, is much greater than Fermi momentum for electron Fermi energy on resonance with the defect, $\varepsilon_{LS} = \varepsilon_F$. This allows us to disregard the first term in (S9) as compared to the second term. After making this approximation, plugging the matrix element $|M|^2$ in (S7) and carrying out standard algebra, we arrive at an expression for the cooling rate which is a product of the defect density of states and a function of the electron and phonon temperatures

$$P_{ep}(\varepsilon) = A(\varepsilon)\left((k_B T_e)^5 - (k_B T_p)^5\right), \quad A(\varepsilon) = \frac{48\zeta(5)}{\pi^2}\,\frac{D^2}{\hbar^3 \rho s^4}\, \frac{(\pi/2)^4 \varepsilon^2 \ln^2\!\left(\frac{W}{vk_p}\right)}{[(\varepsilon\ln(W/|\varepsilon|)+\delta)^2+(\pi/2)^2\varepsilon^2]^2}, \quad \text{(S10)}$$

where we used the relation between the electron-phonon coupling constant and graphene deformation potential $g^2 = D/2\rho s^2$ with $D \approx 50$ eV and correspondingly $\frac{D^2}{\hbar^2 \rho s^2 v^2} = 1.86 \times 10^{19}$ Joule$^{-1}$. The quantity $P_{ep}(\varepsilon)$ vanishes in equilibrium as a result of detailed balance, but is nonzero when the system is driven out of equilibrium. For numerical estimates we rewrite Eq. (S10) scaling energy by 1 meV and temperature by 1 K:

$$P_{ep}(\varepsilon) = \frac{\varepsilon^2 \ln^2\!\left(\frac{W}{vk}\right)(T_e^5 - T_p^5)\,[1\,\text{K}]^{-5}\,[1\,\text{meV}]^2}{[(\varepsilon\ln(W/|\varepsilon|)+\delta)^2+(\pi/2)^2\varepsilon^2]^2} \times 260\,\text{fW}, \quad \text{(S11)}$$

Figure S5 shows $P_{ep}(\varepsilon)$ derived from Eq. (S11) for several indicated values of $\varepsilon_{LS}$ along with the experimental $\delta T(\varepsilon)$. The cooling power spectrum $P_{ep}(\varepsilon)$ exhibits resonance behavior with a sharp peak at $\varepsilon = \varepsilon_{LS}$ consistent with the experimental findings. Using for illustration $T_p = 4.2$ K, $\Delta T = 1$ K, and $\ln(W/vk_p) \approx \ln(270) \approx 5.6$, the dissipated power at the defect is on the order of 6 pW at resonant conditions (Fig. S5) comparable to the experimental evaluations in Section 5.

Using Eq. (S10) and a definition for the electron-phonon heat conductivity of a defect as $P_{ep}(\varepsilon) = \kappa_{ep}(\varepsilon, T_e, T_p)\Delta T$ we obtain:

$$\kappa_{ep}(\varepsilon, T_e, T_p) = A(\varepsilon)\frac{T_e^5 - T_p^5}{\Delta T} \approx 5 T_e^4 A(\varepsilon) \quad \text{(S12)}$$

which is proportional to the cooling-active part of the single-particle density of states. This result elucidates the meaning of "electron-phonon cooling power spectrum", predicting that the cooling power exhibits a resonance behavior as a function of $\varepsilon$, which is due to an enhancement of phonon emission by trapping of an electron on the quasibound state.

Factorization of the dissipated power $P_{ep}(\varepsilon)$ into a part due to the scatterer that mediates cooling and a part due to hot electrons impinging on the scatterer is similar to that underlying the notion of scattering cross-section introduced in the analysis of scattering probes to separate the properties of the target and the source. We also emphasize that the cooling-active part of the single-particle density of states is distinct from the total DOS measured by STM probes, since STM probes the density of all electronic states exposed at the surface whereas our technique selects only the states that mediate electron-lattice cooling.

In the case (ii), where $k_B T, eV \gg \gamma$, the scatterer energy dependence can be treated as a delta function, which gives a simple expression

$$P_{ep}(\varepsilon) = \int_0^\infty d\omega \left(N_{\omega,e} - N_{\omega,p}\right)\left(n_{\varepsilon_{LS}-\omega} - n_{\varepsilon_{LS}+\omega}\right)\frac{g^2\omega^2 v^2(\mu)v}{\gamma a} \tag{S13}$$

where $N_{\omega,e}$ and $N_{\omega,p}$ are Bose functions evaluated at temperatures $T_e$ and $T_p$, respectively, and $n_\omega$ is electron Fermi distribution. The result (S13) describes a peak of width $T_e$ or $T_p$, whichever is greater. In our experiment the width of the resonance is on the order of 10 meV whereas electron temperature inferred from Joule heat dissipated at the voltage bias values used in our measurement is a few times smaller. We therefore expect the regime (i) to provide a better description of the measurement results than the regime (ii).

## 5. Resonant states as a dominant source of dissipation in clean graphene

### 5.1. Thermal imaging at the flat-band condition

In order to understand the significance of the cooling pathway due to resonant LSs and to quantify its overall contribution to dissipation we performed a set of measurements on a second sample (Sample 2) at a flat-band condition that did not involve the "on" and "off" switching of the atomic sources during measurement. The sample was patterned into a rectangular chamber with four constrictions (Fig. 4F), and a tSOT of 48 nm effective diameter (~100 nm outer diameter) was used. A *dc* current $I_{dc} = 1$ µA rms chopped at 35.5 Hz was applied between two of the constrictions as indicated by the arrows in Fig. 4F. In order to reveal the thermal landscape unperturbed by the tip electrostatic potential we acquired $\delta T(x,y)$ images at a flat-band condition $V_{tg} = V_{tg}^{FB}$ as shown in Figs. 4G. The striking observation, which is prominent in Fig. 4G, was $\delta T(x,y)$ enhanced along the edges of the rectangular chamber. Such behavior is observed at all the positive and negative values of $V_{bg}$ (Figs. 4G, S6A, S8A, and S9A) including the charge neutrality point (Fig. S9A), providing strong evidence that the dissipation occurs predominantly at the graphene edges. The reasoning leading to this conclusion is described below step by step. We note that since the flat-band conditions $V_{tg}^{FB}$ depend on the position of the Fermi energy in graphene and hence on $V_{bg}$ values, the flat-band value $V_{tg} = V_{tg}^{FB}(V_{bg})$ was determined for each $V_{bg}$ separately.

### 5.2. Numerical simulations

In order to understand what sources of dissipation could lead to $\delta T(x,y)$ enhanced along the edges, we performed COMSOL numerical simulations of diffusive heat flow using the Sample 2 shape and the estimated values of 2000 W/m/K and 0.2 W/m/K for the heat conductivities of graphene and $SiO_2$ substrate at 4.2 K respectively (our qualitative conclusions weakly depend on these values). The thickness of graphene in the simulation was taken as 50 nm for the reasons of practical convenience, electric and thermal conductances were scaled by the physical thickness. Figure S6 presents the calculated surface temperature distributions for four cases that describe four different cooling regimes:

*i) Joule heating.* A current of 1 µA is applied to the sample through two constrictions on the right as in the experimental configuration. Self-consistent current distribution and the corresponding local dissipation and heat diffusion were calculated numerically using a uniform sheet resistance value of 180 Ω/□ in graphene. The resulting calculated surface temperature distribution is presented in Fig. S6C. This case yields a temperature map in agreement with expectations for the diffusive electron transport regime. The elevated temperature in the constrictions results from their significantly higher resistance, which also gives rise to the overall temperature gradient from right to left.

*ii) Uniform bulk-dominated dissipation.* Since the ballistic electron transport cannot be simulated in COMSOL we study this regime by assuming for simplicity a uniform heat dissipation density of 80 pW/µm² in the entire area of the graphene heterostructure, which amounts to total heat power of 1.8 nW, a value equal to that in case (*i*). Figure S6D shows the resulting temperature distribution which agrees with the behavior expected for ballistic electron transport with a weak and homogeneous inelastic electron-phonon scattering in the bulk. This case represents the behavior in the bulk of the rectangular chamber ignoring enhanced heating in the constrictions due to the higher current density. The essential observation in cases *(i)* and *(ii)* is that the temperature profile $\delta T(x, y)$ through the bulk of the sample has a maximum in the central region as shown by the red and cyan lines in Fig. S6B. A similar behavior is also expected in the case of hydrodynamic electron flow in the presence of bulk-dominated inelastic electron-phonon scattering. The general reason for temperature to peak away from the sample edges is that in the case of bulk dissipation the edges of the sample will always be colder than the sample interior due to a more efficient cooling to the substrate along the sample perimeter.

*iii) Edge-dominated dissipation with diffusive heat propagation.* We study edge-dominated dissipation by carrying out simulations similar to case *(ii)* in which the same total power of 1.8 nW is dissipated at discrete point defects positioned along the edges of the sample (spaced 250 nm apart, as marked by blue points in Fig. S6E). Dissipated power was 12.7 pW per defect, and no dissipation in the bulk. The resulting temperature distribution in Fig. S6E exhibits a pronounced minimum of $\delta T(x, y)$ in the center of the sample as seen by the blue profile in Fig. S6B. This is in stark contrast to the two bulk-dominated dissipation cases *(i)* and *(ii)* characterized by temperature peaks at the sample center. We checked that the temperature minimum is rapidly washed out upon adding a small bulk dissipation. The semblance to the pronounced minimum at the center of the sample observed in experiment (green curve in Fig. S6B) suggests that dissipation in clean encapsulated graphene is dominated by the edge defects and any bulk dissipation, if present, is substantially weaker.

*iv) Edge-dominated dissipation with ballistic phonon propagation.* Interestingly, our measurement results show a minimum in $\delta T(x, y)$ that is even more pronounced than the simulated case of a pure edge dissipation (see temperature profile displayed in Fig. S6B). We speculate that this finding points to the ballistic propagation of emitted phonons. Indeed, in the case of heat propagation dominated by the 2D heat flow in graphene, solution of diffusive 2D heat equations results in universal $-\log r$ temperature decay away from a point source. In contrast, in the 2D ballistic phonon regime the resulting phonon density decays much faster, as $1/r$ from the point source, giving rise to steeper temperature decay. Physically, in clean samples at low temperatures phonons are expected to propagate ballistically on our relevant length scales. In this case the measured $\delta T(x, y)$ should be thought of as a map of the excess power density of the vibrational modes given by the local density of the excess phonons and by their energy. In order to simulate the ballistic phonon regime we use a simple model temperature distribution of $\delta T(r) = \delta T_0 (1 + (r/\ell)^2)^{-1/2}$ with $\delta T_0 = 100$ µK imposed around each defect (shown by the dashed $\delta T_{on}(x)$ line in Fig. 3B). Here $\ell$ is a characteristic length scale for the cut-off of the $1/r$ divergence determined by the diameter of the tSOT and the scanning height above the graphene, which we took as $\ell = 100$ nm. Figure S6F shows the resulting $\delta T(x, y)$ given by superposition of such point sources positioned at the same locations as in Fig. S6E. As illustrated in Fig. S6B, this simple model gives a good agreement with the experimental results, providing strong support for the dissipation mechanism that is dominated by inelastic electron scattering by resonant LSs at the clean graphene edge.

To simplify the analysis, in Fig. S6F we have assigned the same power dissipation to all the defects. In reality, however, the LSs in the current-carrying constrictions will dissipate more power due to the larger local current density resulting in higher collision rate of the electrons with the LSs. This will result in the higher intensity of the rings in the constrictions and in their proximity as indeed clearly seen in Figs. 4G, S6A, S8A, and S9A.

### 5.3. The density of cooling-active LSs at graphene edge

Figures 4G, S6A, S8A, and S9A demonstrate that edge-dominated dissipation is present at all values of $V_{bg}$. Since each defect creates a dissipative resonant state at a specific energy, this observation implies that different defects contribute to dissipation at different $V_{bg}$. This is illustrated by a wide spread of resonance energies in Fig. 4 in the main text. Signatures of a multitude of different LSs can be discerned also from the measurements under the flat band condition. Indeed a closer inspection of Figs. 4G, S6A, S8A, and S9A shows inhomogeneity in $\delta T(x, y)$ along sample edges that changes upon varying $V_{bg}$. In Fig. 4 our numerical algorithm detected 135 LSs along $L = 3.5$ μm along graphene edge in the energy window of ~260 meV (SM section 10). Taking the spectral width of each LS to be about 13 meV (Fig. 2 and SM Section 9), this implies that at any given $V_{bg}$ there are about 7 out of 135 defects which are cooling-active. This corresponds to an average separation of 500 nm for the active defects. Since our algorithm detects only a fraction of the LSs, the actual density of the cooling-active LSs is probably somewhat greater. In simulations shown in Figs. S6E,F we used point heat sources spaced by 250 nm.

### 5.4. Resolving localized states by $V_{bg}^{ac}$ modulation

Since the temperature fields created by nearby LSs overlap, their individual contributions cannot be readily resolved in the $\delta T(x, y)$ images of Figs. 4G and S6A at a flat-band condition. Upon variation in $V_{bg}$ some LSs move out of resonance, reducing their contribution to dissipation while at the same time other LSs are shifted into resonance forming new point sources of dissipation. By applying an *ac* voltage $V_{bg}^{ac} = 0.1$ V to the back gate in addition to the *dc* $V_{bg}^{dc}$ the details of this process can be resolved by imaging the corresponding *ac* change $T_{ac}^{bg}$ in the local temperature. Figure S7 shows two examples of such $T_{ac}^{bg}(x, y)$ images acquired at $V_{bg}^{dc} = -8$ V and –6.7 V. The bright and dark spots that appear at random locations along the edges of graphene reflect the LSs that are correspondingly shifted in and out of resonance by the $V_{bg}^{ac}$ change in $V_{bg}$. Note that the dark spots correspond to negative $T_{ac}^{bg}(x, y)$, namely in these regions the dissipation decreases as $V_{bg}$ increases, while in the bright spots $\delta T(x, y)$ increases with $V_{bg}$ giving rise to positive $T_{ac}^{bg}(x, y)$. This picture would look quite differently if the microscopic dissipation sources were independent of $V_{bg}$. In this case, instead of strong bright and dark spots we would see only a weak and smooth constant-sign background $T_{ac}^{bg}(x, y)$ originating from the gate dependence of graphene resistivity.

### 5.5. Resolving individual LSs by tip gating

The above $V_{bg}^{ac}$ method of imaging the resonant LSs is somewhat limited since it reveals only the states that are close to resonance at a given $V_{bg}$, which makes resolving individual LSs that are located close to each other a challenging task. However, by varying the tip potential $V_{tg}$ and moving away from the flat-band condition, a wide range of LSs can be individually resolved with very high spatial resolution. This enhanced functionality arises from the fact that instead of imaging the net $\delta T(x, y)$ which is a sum of contributions due to all active LSs contributing to dissipation at a particular $V_{bg}$, the local potential induced by the tip can be used for selective and position dependent imaging of individual LSs. This is done by switching individual LSs in and out of resonance and measuring the corresponding change in the local temperature $T_{ac}^{TF}(x, y)$ due to modulation of a single resonant state by the vibrating tip. The striking difference between $\delta T(x, y)$ at a flat-band condition and the local temperature change $T_{ac}^{TF}(x, y)$ in the presence of tip potential is illustrated in

Fig. S8. While Fig. S8A clearly shows that the dissipation occurs predominantly along the graphene edges, the presence of individual LSs cannot be discerned. Figure S8B, in contrast, reveals the high density of resonant LSs at graphene edges with a wide-spread energy level distribution giving rise to numerous rings of different diameters. The wealth of information provided by the $T_{ac}^{TF}(x,y)$ imaging is the reason it is mainly the data obtained by this method that are presented in the main text.

The amount of detail obtained by this method is illustrated in Figure S8 which provides an interesting insight into the origin of dissipation enhancement in the current-carrying constrictions. In narrow constrictions both the current density and the edge-to-bulk ratio are high. As a result the rates for electron collision with the LSs and for the corresponding phonon emission are much higher than in the chamber, resulting in enhanced heating in the two constrictions on the right that serve as current source and drain. The high rate of phonon emission in the constrictions contributes to the overall right to left gradient in $\delta T(x,y)$ within the chamber seen in Fig. S8A. This gradient is significantly enhanced near CNP as presented in Fig. S9 and observed as the bright horizontal streaks in Figs. S11D and S11F around $V_{bg} = 0$. Interestingly, the ballistic nature of electron transport in our clean graphene causes dissipation also within the two floating constrictions on the left where no current flows. We interpret the ring structures seen in the floating constrictions in Fig. S8B as being due to the hot carriers that travel ballistically across the chamber, enter the constrictions and dissipate their energy while dwelling inside the constrictions.

### 5.6. Current dependence

The observed overall temperature increase in the sample arises predominantly from the phonons that are emitted by the resonant LSs at graphene edges and propagate throughout the sample rather than being generated in the bulk of graphene. In order to study the current dependence of the overall sample temperature we carried out thermal imaging at flat-band condition as presented in Fig. S9. Temperature profile across the sample along the dashed line in Fig. S9A was then acquired at different values of the applied current $I_{dc}$. Examples of the resulting $\delta T$ profiles at different currents are presented in Figure S9B showing the temperature increase with the current $I_{dc}$. Figure S9C presents the average temperature $\langle \delta T \rangle$ along the profiles vs. the current, showing $\langle \delta T \rangle \propto I_{dc}^2$ behavior over a wide range of currents down to our lowest current of 20 nA. Such low currents are well comparable to the currents typically employed in transport studies thus indicating the relevance of the observed dissipation mechanisms to the transport properties of graphene.

### 6. Simulations of the electrostatic interaction between the tSOT and the sample

The powerful spectroscopic capabilities of the tSOT nanothermometry arise from the local electrostatic interaction between the tip and sample. In this section we derive this interaction and provide some simulations by numerical solution of Laplace's equation for the potential $\varphi(r,z)$ in an axial-symmetric geometry.

Our simulated space consisted of graphene residing in $z = 0$ plane while the tSOT was modeled for simplicity as a disc with outer diameter $d_0$ and thickness $t$ located at a height $z_{tip}$ above the graphene. A fixed voltage boundary condition of $V_{tg}$ was applied to the disc and a fixed voltage boundary condition of $V_{bg}$ was applied at $z = -d_{bg}$ to account for the back gate. The entire space above and below the graphene was taken to be homogeneous with dielectric constant $\epsilon$ to simplify the numerics. Zero Neumann boundary conditions were used at large $r$ and $z$. The surface charge density $\sigma(r)$ in the graphene at $z = 0$ was solved self-consistently as follows. The tip potential induces a local band bending $E_D(r)$ of the graphene Dirac energy and changes the local chemical potential $\mu(r) = E_F - E_D(r)$, while the electro-chemical potential $E_F$ remains constant throughout the graphene and fixed to $E_F = 0$. Unlike in an ideal metal, a finite in-plane field can be present in graphene. Equilibrium conditions require that a test charge experiences a zero net in-plane force. This dictates that the in-plane field is balanced by the gradient of the chemical potential, which results in a boundary condition $E_D(r) = -e\varphi(r, z = 0) + E_{D,\infty}$, where $E_{D,\infty}$ is the global band bending due to the back-gate voltage $V_{bg}$. Considering the Dirac density of states $D(\varepsilon) = \partial n/\partial \varepsilon = 2\varepsilon(\sqrt{\pi}\hbar v_F)^{-2}$ (where $\varepsilon = |E_F - E_D|$) and the geometrical capacitance between the two parallel plates (graphene and back gate) it is straightforward to derive at analytic expression $E_{D,\infty} = -sign(V_{bg})\sqrt{\pi}\hbar v_F\sqrt{n_\infty}$, where:

$$n_\infty = \left(\sqrt{\left(\frac{\epsilon\epsilon_0\sqrt{\pi}\hbar v_F}{2e^2 d_{bg}}\right)^2 + \frac{\epsilon\epsilon_0}{e\cdot d_{bg}}|V_{bg}|} - \frac{\epsilon\epsilon_0\sqrt{\pi}\hbar v_F}{2e^2 d_{bg}}\right)^2$$

On the other hand, $\sigma(r)$ is determined by $E_D(r)$ through $D(\varepsilon)$, resulting in $\sigma(r)$ that is given by the relation $E_D(r) - E_F = \text{sign}(\sigma(r))\sqrt{\pi/e}\hbar v_F\sqrt{|\sigma(r)|}$, resulting in a closed set of equations that can be self-consistently solved.

In the simulation results of Fig. S10 the following parameter values were used: outer tip diameter $d_0 = 50$ nm, tip thickness $t = 5$ nm, $\epsilon = 3$, $v_F = 10^6$ m/s, $d_{bg}$ was set to 125 nm to match the experimentally derived $C_{bg}$ (see Section 1 above), and $z_{tip}$ was set to 57 nm to attain comparable $C_{tg}$ and $C_{bg}$ capacitances as in the experiment. The graphene band bending $E_D(x)$ is shown in Fig. S10A for tip potential $V_{tg} = -1.5$ V for various back gate voltages $V_{bg}$. The $E_D(x)$ curve in Fig. 3 and Movie S3 was derived for $V_{tg} = -3$ V and $V_{bg} = 0.25$ V. The strongest tip-induced band bending occurs when $E_D(x)$ crosses zero where the graphene is incompressible. At high values of $V_{bg}$ the bending is reduced due to a more effective screening. Figure S10B shows similar results for a constant $V_{bg} = 1$ V and different values of $V_{tg}$. From these results we can calculate the evolution of the radius $R$ of the dissipation ring assuming that the LS is pinned to the Dirac point, $E_{LS} = E_D(x)$, where $x$ is the tip distance to the defect, and the carriers are injected close to the Fermi energy $E_e \cong E_F = 0$. Figure S10C shows the resonant dissipation traces as a function of $V_{tg}$ for various values of $V_{bg}$ akin the experimental results in Figs. S11A-C. The points along the traces indicate the lateral displacement $x$ of the tSOT from the defect at which resonant conditions for inelastic scattering are met for various values of $V_{tg}$ and $V_{bg}$. For example, for $V_{bg} = 1$ V, as described in Fig. S10B, the resonant conditions occur only for negative $V_{tg} < -1$ V for which $E_D(x)$ crosses $E_F = 0$. The distance $2R$ between the crossing points grows with decreasing $V_{tg}$ resulting in the resonance trace marked by an arrow in Fig. S10C. Figure S10D presents the resonant bell-shaped traces as a function of $V_{bg}$ for various values of $V_{tg}$ describing the experimental results in Figs. 4A-C.

## 7. Search for additional energy levels of the localized state

In this section we provide additional information on the spectroscopic analysis of bulk defects and explore the widest energy range for searching for additional possible energy levels of the LS. By repeatedly scanning the tSOT along the line crossing defect 'C' and incrementing $V_{tg}$ we derive the resonant traces describing the evolution of the radius of the dissipation ring $R(V_{tg})$ as shown in Figs. S11A-C. These traces correspond to the numerical results presented in Fig. S10C. In the case of $V_{bg} = 2.5$ V (Fig. S11C), the concave trace reflects the situation exemplified in Fig. 3 where in absence of tip potential, $E_{LS}$ resides below $E_F$ and hence the LS is occupied and does not dissipate until a negative $V_{tg}$ brings it into resonance with $E_e$. For $V_{bg} = -1$ V (Fig. S11A), the opposite situation occurs in which the empty state does not dissipate because it is well above $E_e$ and a positive $V_{tg}$ is required for bringing it into resonance resulting in a convex trace. The transition from convex to concave traces upon varying $V_{bg}$ is presented in Fig. S11B and Movie S4.

Alternatively, the tSOT can be scanned through the defect while incrementing $V_{bg}$ at constant $V_{tg}$ (Figs. 4A-C and Movie S5). In this case the traces are bell-shaped, in which $R$ diverges when $V_{bg} = V_{bg}^{LS}$ as expected (Fig. S10D). Figures S11D and S11F show the bell-shaped resonance traces at $V_{tg} = 10$ V and $V_{tg} = -10$ V over the entire range of $V_{bg} = \pm 10$ V. Traces at various intermediate values of $V_{tg}$ are presented in Fig. S11E. These data are summarized in Fig. S11G showing the ring radius $R$ in the $V_{bg}$-$V_{tg}$ plane. In the white areas no dissipation is present since $E_{LS}$ is far from resonance. In the colored regions inelastic scattering occurs depending on the tip position. The transition line (dark blue) maps the resonant conditions when the tip is positioned directly above the defect ($R = 0$) as in Fig. 2D, while the red color reflects the conditions for diverging $R$. The black dot describes the bare resonance point $V_{bg}^{LS}, V_{tg}^{FB}$ at which $V_{tg}$ corresponds to flat-band conditions of the tip and $V_{bg}$ aligns $E_{LS}$ with $E_e \cong E_F$ in absence of tip potential. This is the point at which the resonant traces switch their polarity and $V_{bg}^{LS}$ describes the position of the LS energy level $E_{LS}$.

The slope of the $R = 0$ resonance line in Fig. S11G (dark blue) describes the ratio between the back gate and the tip capacitances $C_{bg}/C_{tg}$ that is affected by the tip height $h$ as described in Section 6 above. With $C_{bg} = 1.25 \cdot 10^{11}$ 1/cm²V attained from the magneto-transport data of Section 1, we derive $C_{tg} = 1.3 \cdot 10^{11}$ 1/cm²V at $h = 12$ nm.

The fact that only a single resonance trace is visible in Figs. S11D and S11F shows that the LS has just one energy level in this range of $\pm V_{tg}^{max}$ and $\pm V_{bg}^{max}$. This translates into a range of carrier concentrations of $\pm(C_{tg}V_{tg}^{max} + C_{bg}V_{bg}^{max}) = \pm 2.6 \cdot 10^{12}$ cm⁻² and corresponding energy span of $\pm 186$ meV. These results show that the LS has just a single energy level $E_{LS}$ near the Dirac point with no additional levels in the range of at least $\Delta E = \pm 186$ meV from $E_D$. Considering the LS as a quantum dot, the single electron changing energy and hence the corresponding energy level separation is given by $\Delta E \cong e^2/(4\pi\varepsilon_{hBN}\varepsilon_0 d_{LS})$ where $d_{LS}$ is the diameter of the LS and $\varepsilon_{hBN} \cong 3$ is the dielectric constant of hBN. Taking the lower bound estimate of $\Delta E \geq 186$ meV we attain the maximal effective size of the LS $d_{LS} \leq 2.4$ nm.

## 8. Derivation of the local CNP and of the energy level $E_{LS}$ of bulk localized state

The LS resonance line in Fig. 2D provides a valuable tool for determining the local charge neutrality point (CNP) in terms of $V_{bg}^{CNP}$ and the energy level $E_{LS}$ of the LS in terms of $V_{bg}^{LS}$. To describe the derivation method we carry out numerical simulations of the resonance line, which is the critical line at which the dissipation ring nucleates from $R = 0$.

We first assume that the LS is pinned to the Dirac point, $\varepsilon_{LS} = E_{LS} - E_D = 0$. In this case the resonance curve is defined by the geometric place of all the $V_{bg}, V_{tg}$ pairs for which the $E_D(x)$ profile at the tip location $x = 0$ is tangent to the Fermi energy $E_F = 0$ as shown in Fig. S12A, such that the impinging electrons with energy $E_e \cong E_F$ are at resonance with the LS. The resulting $R = 0$ resonance curves are shown in Fig. S12B for various heights $z_{tip}$ of the tip above the graphene. The slope of the curves away from the CNP reflects the ratio between the effective capacitances $C_{bg}/C_{tg}$ of the back and tip gating and is thus affected by $z_{tip}$. The intersection point of the curves is given by the flat-band condition of the tip $V_{tg}^{FB}$ which depends on its work function and is taken to be zero in the simulations. At $V_{tg}^{FB}$ the tip has no effect on the sample, $E_D(x)$ is constant, and the resonance occurs at $V_{bg} = V_{bg}^{LS}$ independent of the tip height. The crossing point thus determines the energy of the LS in terms of $V_{bg}^{LS}$.

The resonance curves are not straight and show a kink which is clearly resolved by plotting the derivatives of the curves (Fig. S12C). This kink is the result of the quantum capacitance and reflects the change in effectiveness of graphene in screening the tip potential at different $V_{bg}$. The kink occurs at the charge neutrality point $V_{bg}^{CNP}$, which is taken to be zero in the simulations, and is defined by the location of the minimum of the $dV_{tg}/dV_{bg}$ curves (Fig. S12C). In the $\varepsilon_{LS} = 0$ case, $V_{bg}^{LS} = V_{bg}^{CNP} = 0$ and hence the crossing point of the resonance curves coincides with the kink location, however, in the general case these two features are separated as demonstrated in Figs. S12D-F for $\varepsilon_{LS} = 30$ meV. Here the crossing point of the curves is determined by the top-gate flat-band condition $V_{tg}^{FB} = 0$ at which $V_{bg} = V_{bg}^{LS}$, while the kink is fixed by the back gate CNP, $V_{bg}^{CNP} = 0$ (Figs. S12E-F).

In the experimental resonance lines of defect 'C' in Fig. S13D (and their derivative in Fig. S13E) the two features are well resolved with resulting $V_{bg}^{LS} \cong 200$ mV and $V_{bg}^{CNP} \cong 480$ mV. From these two values we derive the energy level of the LS to be

$$\varepsilon_{LS} = E_{LS} - E_D = sign(V_{bg}^{LS} - V_{bg}^{CNP})\sqrt{\pi/e}\hbar v_F\sqrt{C_{bg}|V_{bg}^{LS} - V_{bg}^{CNP}|} \cong -22 \text{ meV}.$$

Figures S13A-C show the corresponding numerical simulations for the case of $\varepsilon_{LS} = -20$ meV reproducing the experimental features. In order for the LS to be at resonance with the electrons at $E_F$ the tip potential has to induce $E_D(x)$ profile such that $E_D(0) - E_F = 20$ meV (Fig. S13A).

Theoretical calculations and STM studies (*13*, *14*, *16*, *17*) show that carbon vacancies and monovalent adatoms form LSs close to the Dirac point consistent with our findings. In particular, hydrogen atoms on graphene were shown (*13*, *14*) to give rise to sharply localized state at 30 meV below the Dirac point. Our derived $\varepsilon_{LS} \cong -22$ meV thus raises the possibility that the observed resonant inelastic scattering in the bulk of graphene may arise from individual hydrogen adatoms.

Note that in graphene heterostructures the CNP is known to be hysteretic with respect to $V_{bg}$. Accordingly, $V_{bg}^{CNP}$ and $V_{bg}^{LS}$ are slightly dependent on the span of previously applied $V_{bg}$ and on its sweep direction, giving rise to small variations in the apparent $V_{bg}^{LS}$ in Figs. 4A-C and S11A-C. Since $\varepsilon_{LS}$ is determined only by the difference $V_{bg}^{LS} - V_{bg}^{CNP}$ the above described procedure diminishes the effect of hysteresis by measuring the two quantities in a single continuous $V_{bg}$ sweep.

The bulk defects are extremely rare in high quality encapsulated graphene. Besides the three defects in Sample 1 we found only four additional bulk defects in other samples, while in some of the samples, like Sample 2, we could not resolve any bulk defects. From this limited statistics we can estimate the average areal density of $2 \cdot 10^5$ cm$^{-2}$ of bulk defects in graphene. Such a low density and their atomic nature may indicate that the bulk defects are not formed during the sample fabrication process but rather originate from native defects in the parent graphite with a corresponding concentration of $5 \cdot 10^{-5}$ ppm. The spectral properties of bulk defect 'C' were thoroughly investigated as presented here. The spectral properties of the other six defects were attained by measurements similar to those presented in Movies S4-5. From these assessments we conclude that for all bulk defects the LS energy level resides within at most 50 meV from the Dirac point, pointing to a common chemical or structural origin in form of carbon vacancies or hydrogen adatoms.

## 9. Experimental derivation of the electron-phonon cooling power spectrum

Our thermal spectroscopy allows evaluation of the broadening $\delta E$ of the energy level of the LS and derivation of its electron-phonon cooling power spectrum (CPS), $P_{ep}(\varepsilon)$, as follows. For a LS that is pinned to the Dirac point, $E_{LS} = E_D$, a finite spectral width $\delta E$ of the LS translates into a finite width of the back gate voltage $\delta V_{bg}$, $\delta E = 2\sqrt{\pi/e}\hbar v_F\sqrt{C_{bg}\delta V_{bg}/2}$, over which the dissipation occurs. This broadening translates into a finite width $\delta x = \delta V_{bg}/(dV_{bg}/dx)$ of the dissipation ring as described in Fig. S14. In addition to the intrinsic broadening, however, there is an extrinsic broadening due to a finite energy distribution of the injected energetic carriers and the tSOT modulation $x_{ac}$ by the TF.

In order to assess the extrinsic broadening and to minimize its contribution we performed the measurements presented in Fig. S14. Figure S14A shows the bell-shaped dissipation trace of defect 'B' at $V_{tg} = -10$ V. The back-gate voltage was then fixed to $V_{bg} = 5.3$ V, where the slope of the bell-shaped trace was $dV_{bg}/dx = 47$ mV/nm, as shown in Fig. S14A. Next, we performed line scans through the LS and measured the width of the ring $\delta x$ (Fig. S14D) as a function of the rms amplitude of the tip oscillation $x_{ac}$ as shown in Fig. S14B. The resulting $\delta x$ decreases linearly with decreasing $x_{ac}$, as expected when tip oscillation is larger than the intrinsic width of the ring, and saturates for $x_{ac} \lesssim 1.5$ nm. The average energy of the injected electrons $\varepsilon_e = E_e - E_D$ and the width of their energy distribution, which should be comparable to $\varepsilon_e$, are determined by $R_c$ and the *dc* current $I_{dc}$, $\varepsilon_e = eR_cI_{dc}$, where $R_c = 890$ Ω is the resistance of the constriction at $V_{bg} = 5.3$ V (see Fig. S1). Figure S14C shows that the ring width $\delta x$ decreases with decreasing $I_{dc}$ and saturates for $I_{dc} \lesssim 5$ μA. The corresponding dissipation ring measured in the limit of low tip oscillation of $x_{ac} = 0.5$ nm and low current $I_{dc} = 2$ µA is presented in Fig. S14D (note the $T_{ac}$ scale of just 1 µK). The cross section $T_{ac}(x)$ of the ring (Fig. S14E) shows ring width of $\delta x = 7$ nm.

After minimizing the extrinsic contributions to the broadening, we can directly derive the CPS from the spectroscopic thermal signal, $P_{ep}(\varepsilon) \propto \delta T(\varepsilon)$. To attain this, we integrate the $T_{ac}(x)$ profile of Fig. S14E to derive $\delta T(x) = \int^x dx' T_{ac}(x')/x_{ac}$, and use coordinate transformation to translate the lateral tip position $x$ into the induced energy shift of the LS given by $|\varepsilon| = \sqrt{\pi/e} \hbar v_F \sqrt{C_{bg}|V_{bg}^{LS} - V_{bg}^{CNP} + x\, dV_{bg}/dx|}$, where $dV_{bg}/dx$ is determined from Fig. S14A, the origin of $x$ is defined in Fig. S14E, and $V_{bg}^{LS} - V_{bg}^{CNP}$ was taken from Section 8 above. The resulting $\delta T(\varepsilon)$ is presented in Fig. S14F and describes the CPS of the LS, $P_{ep}(\varepsilon) \propto \delta T(\varepsilon)$, which shows a sharp peak with spectral width of $\delta E = 13$ meV. Since $\varepsilon_e = eR_c I_{dc} = 1.8$ meV $\ll \delta E$, the observed $\delta E$ reflects an upper bound on the intrinsic broadening of the energy level of the LS. The derived $P_{ep}(\varepsilon)$ is presented in Fig. 2F.

## 10. Analysis of localized states along the graphene edges

The numerous LS along the graphene edges were analyzed as described in Fig. S15. The tSOT was scanned along the bottom edge of the graphene sample and the bell-shaped resonance traces were acquired by sweeping the back gate $V_{bg}$ at various $V_{tg}$ (see Movie S6). Figure S15A shows the data at $V_{tg} = -10$ V reproduced here from Fig. 4 for clarity. Figure S15B presents the same data with overlaid fits to the traces (black) of empirical form $V_{bg}(x) = V_{bg}^{LS} - V_{bg}^{CNP} + V_p/(1 + (x - x_i)^2/w^2)$ with four fitting parameters: $V_{bg}^{LS} - V_{bg}^{CNP}$ is the $V_{bg}$ value corresponding to the energy $E_{LS}$ of the LS, $V_p$ describes the height of the bell shape determined by $V_{tg}$, $w$ is the width of the bell shape governed by the scanning height $h$ of the tSOT, and $x_i$ is the location of the LS along the edge.

Our numerical algorithm detected $m = 135$ traces along $L = 3.5$ µm long graphene edge which reflects only a fraction of the total number of LSs. The average density of the identified LSs is $\rho = m/L = 38.6$ µm$^{-1}$ corresponding to average distance of $\langle \Delta x \rangle = 26$ nm. Analysis of the distances $\Delta x_{nn}$ between the nearest neighbors (Fig. S15C) reveals that the probability density of $\Delta x_{nn}$ is described by Poisson distribution $P(\Delta x_{nn}) = \rho e^{-\Delta x_{nn} \rho}$ (black curve) indicating that LSs are distributed randomly, excluding presence of clustering or long-range spatial correlations. Interestingly, Fig. S15E shows $\Delta x_{nn}$ separations of down to below 1 nm putting an upper bound on the spatial extent of the individual LSs and emphasizing the atomic-scale nature of the defects.

Figure S15D presents the $V_{bg}^{LS}$ values of the different LSs that shows a very wide distribution with no apparent correlations. Such a large random variability over very short length-scales is inconsistent with the possibility that the variations in $V_{bg}^{LS}$ result from formation of "puddles" due to long-range substrate potential disorder (*21*) and support the Coulomb repulsion scenario. Moreover, the $V_{bg}^{LS}$ values appear to be uniformly distributed, without visible clustering around specific values that may characterize particular adatom defects. Chemical variability of adatoms is therefore not a likely explanation of the large variability of the energy of the LSs. The difference in $V_{bg}^{LS}$ between the nearest neighbors $\Delta_{nn} V_{bg}^{LS}$ (Fig. S15E) reaches $\pm 15$ V, limited by our accessible $V_{bg}$ span, and seems to be independent of the distance $\Delta x_{nn}$ between the neighbors from sub nm to 100 nm distances. Note, however, that we analyze only a fraction of the defects. Remarkably, we detect neighboring defects that are separated on nm scale while showing variation of up to 10 V in their $V_{bg}^{LS}$ energy. Examples of such pairs of nearest neighbors are marked in color in Fig. S15B. The green traces show an example of five LSs within a segment of 8 nm along the graphene edge with significantly different energies.

**Supplementary Figures**

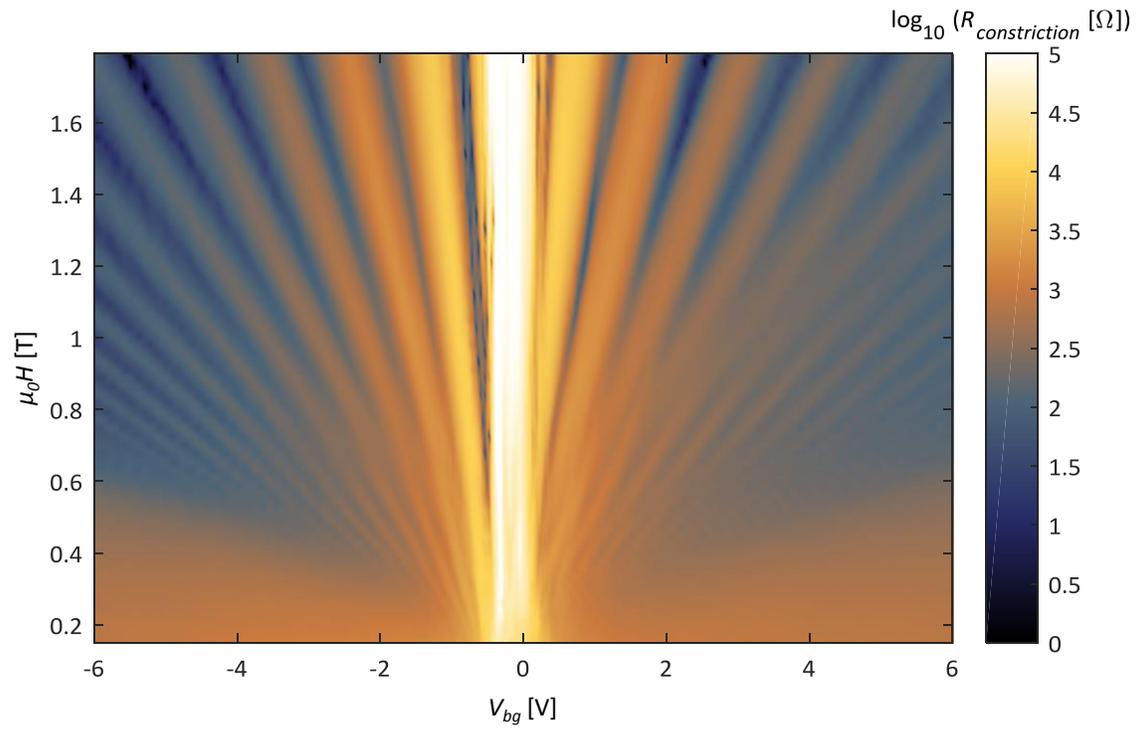

**Figure S1. Magneto-transport characterization of the graphene sample.**

Four-probe measurement of $R_{xx}$ of the top constriction of the sample in Fig. 1B vs. back gate voltage $V_{bg}$ and the perpendicularly applied magnetic field $H$ at $T = 4.2$ K, showing quantum Hall oscillations. A linear fit to the QH minima results in back gate capacitance of $C_{bg} = 2 \cdot 10^{-4}$ F/m².

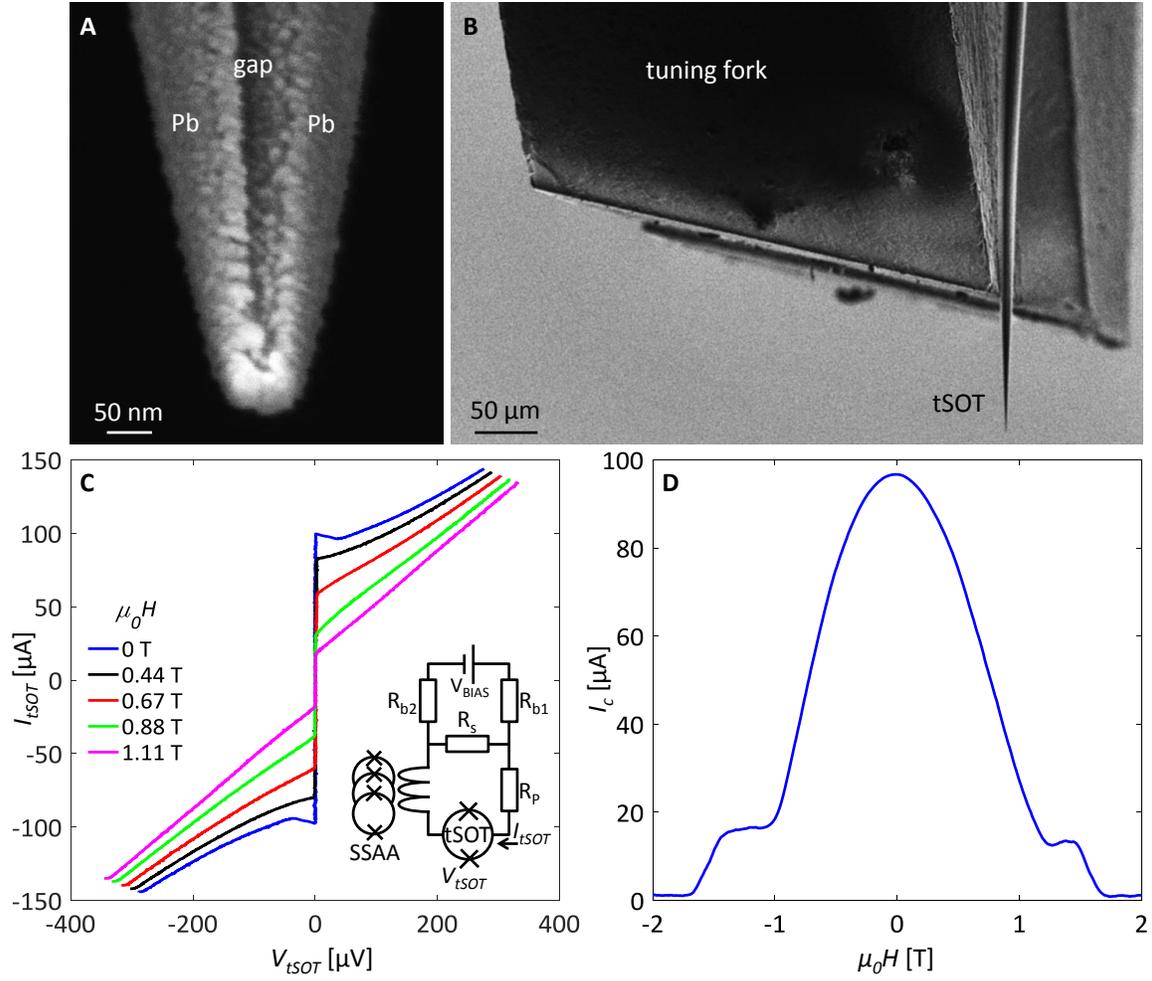

**Figure S2. tSOT characterization.**

(**A**) SEM image of the tSOT used in this work showing two Pb leads connecting to the Pb ring at the apex of the tip and the gap between them. (**B**) SEM image of the tSOT pipette attached to one tine of the quartz tuning fork. (**C**) I-V characteristics of the tSOT at 4.2 K and various applied magnetic fields. The inset shows a simplified measurement circuit which includes the bias source $V_{BIAS}$, bias resistors $R_{b1} = R_{b2} = 11$ kΩ, shunt resistor $R_s = 5.4$ Ω, parasitic resistance $R_p = 0.6$ Ω, and the inductively coupled SSAA. (**D**) Quantum interference pattern of the critical current $I_c(H)$ showing a period of 2.4 T corresponding to effective diameter $d = 33$ nm of the tSOT. The higher lobes of the interference pattern are suppressed by $H_{c2}$ of Pb.

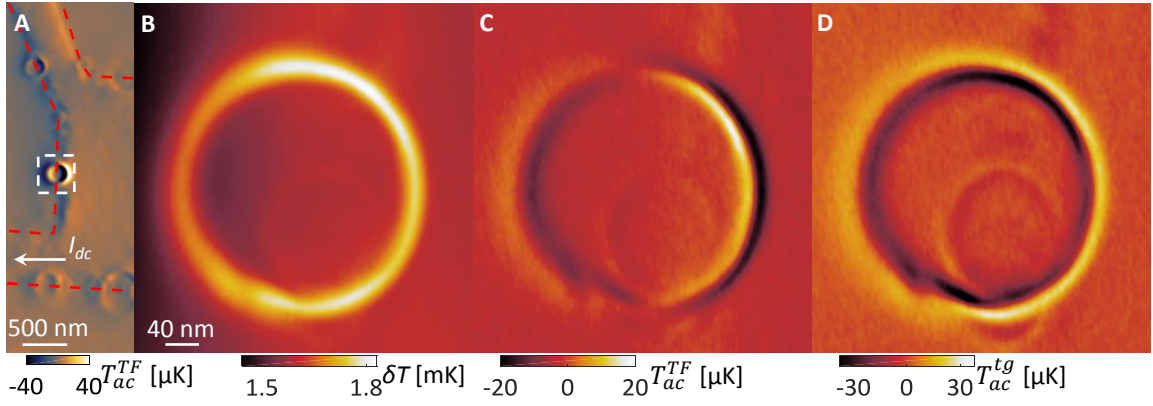

**Figure S3. Demonstration of simultaneous measurement by three thermal imaging methods.**

(**A**) $T_{ac}^{TF}(x,y)$ image of the left part of Sample 2 revealing the dissipation rings along the edges of hBN/graphene/hBN heterostructure (dashed). The arrow indicates the current $I_{dc}$ (chopped at 35.5 Hz) that flows from the right side of the sample into the constriction. (**B**) Zoomed-in thermal image $\delta T(x,y)$ of a dissipation ring marked by dashed square in (**A**). (**C**) Simultaneously measured ac thermal image $T_{ac}^{TF}(x,y)$ due to $x_{ac}$ tip oscillation by the TF, $T_{ac}^{TF}(x,y) = x_{ac} d\delta T(x,y)/dx$. (**D**) The *ac* thermal image $T_{ac}^{tg}(x,y)$ due to $V_{tg}^{ac}$ tip gate modulation, $T_{ac}^{tg}(x,y) = V_{tg}^{ac} dT(x,y)/dV_{tg}$. Scan parameters: (**A**) Scan area $1.4 \times 3.9$ µm², pixel size 13 nm, scan-speed 200 ms/pixel, $h = 20$ nm, $V_{bg} = -2$ V, $V_{tg} = 2$ V, $I_{dc} = 2$ µA rms chopped at 35.5 Hz, $x_{ac} = 1.4$ nm. (**B-D**) Scan area $400 \times 410$ nm², pixel size 3.3 nm, scan-speed 250 ms/pixel, $h = 20$ nm, $V_{bg} = 0$ V, $V_{tg} = 5$ V, $I_{dc} = 1$ µA rms chopped at 35.5 Hz, $x_{ac} = 1.5$ nm at 4.5° to the $x$ axis, $V_{tg}^{ac} = 0.1$ V at 3.06 kHz.

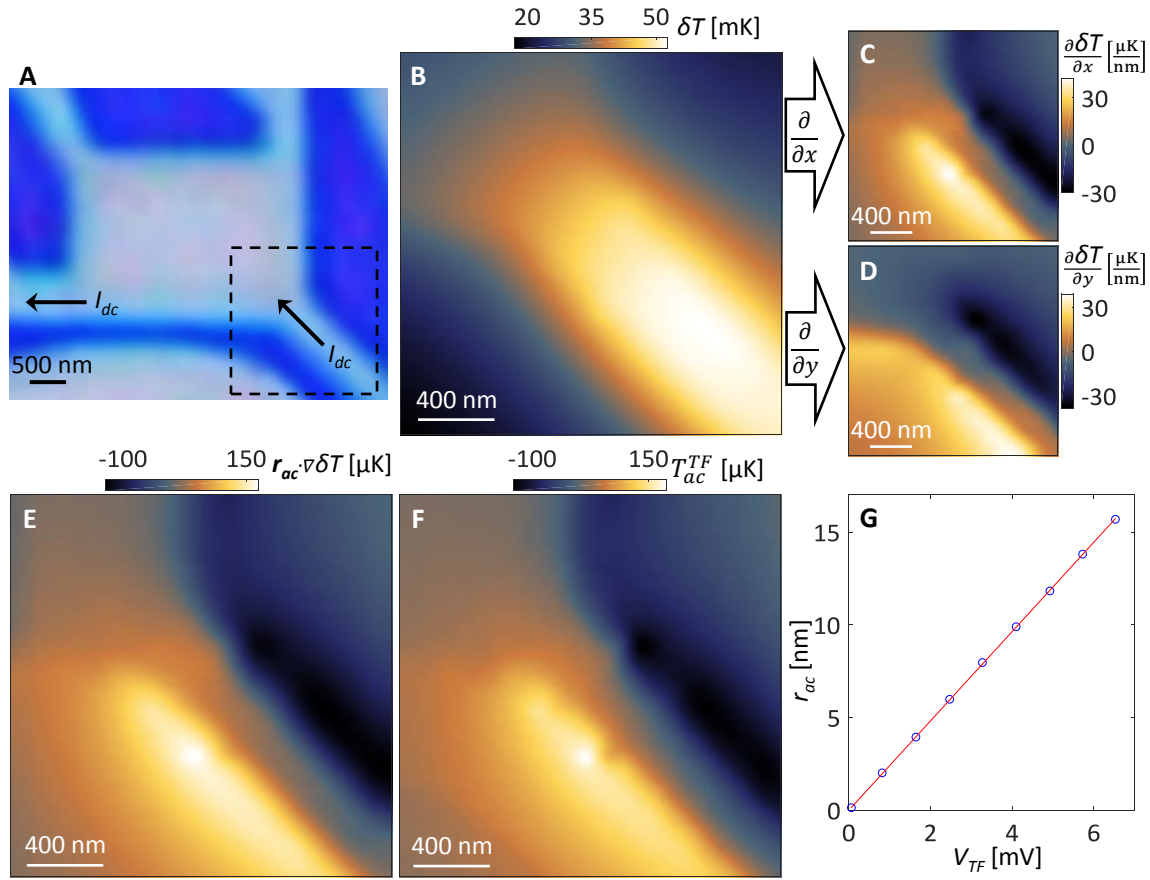

**Figure S4. Measurement of the tSOT vibration amplitude.**

(**A**) Optical image of Sample 2 showing the etched hBN/graphene/hBN heterostructure (bright) on $SiO_2$ substrate (dark). A *dc* current $I_{dc} = 4$ μA rms chopped at 35.5 Hz was applied as indicated by the arrows. (**B**) Thermal $\delta T(x,y)$ image of the constriction region marked by the dashed square in (**A**). (**C,D**) Numerical derivatives $\partial \delta T/\partial x$ and $\partial \delta T/\partial y$ of image (**B**). (**E**) Best fit linear combination of (**C**) and (**D**), $x_{ac}\, \partial \delta T(x,y)/\partial x + y_{ac}\, \partial \delta T(x,y)/\partial y$, to the measured $T_{ac}^{TF}(x,y)$ image (**F**) providing the tSOT vibration of $x_{ac} = 4.05$ nm and $y_{ac} = 0.31$ nm. (**F**) $T_{ac}^{TF}(x,y)$ measured simultaneously with (**B**) at TF excitation $V_{TF} = 1.66$ mV. (**G**) The measured rms tSOT vibration amplitude $r_{ac} = \sqrt{x_{ac}^2 + y_{ac}^2}$ vs. $V_{TF}$ (blue dots) and a linear fit (red) with slope of 2.41 nm/mV. Scan parameters of (**B,F**): Scan area 2×2 μm², pixel size 20 nm, scan-speed 80 ms/pixel, $h = 20$ nm, $V_{bg} = -2$ V, $V_{tg} = -0.53$ V corresponding to $V_{tg}^{FB}$ at this $V_{bg}$.

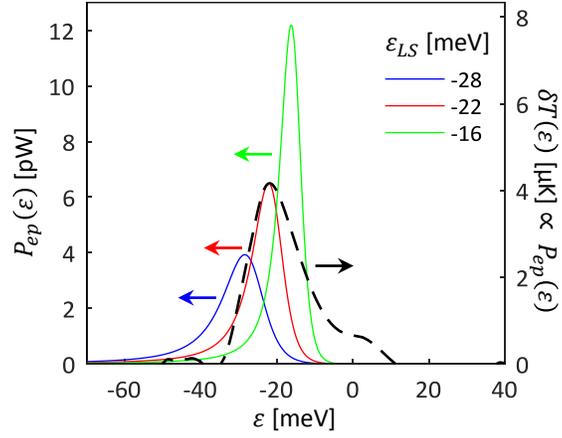

**Figure S5. Electron-phonon cooling power spectrum of a single defect.**

Cooling power spectrum of a quasi-bound state, $P_{ep}(\varepsilon)$, calculated from Eq. (S11) for three indicated values of $\varepsilon_{LS}$ (left axis) and the experimental $\delta T(\varepsilon) \propto P_{ep}(\varepsilon)$ of defect 'C' (right axis) as a function of the Fermi energy $\varepsilon$. Parameters of Eq. (S11): $T_p$ =4.2 K, $T_e$ =5.2 K, $\ln(W/vk_p)$=5.6.

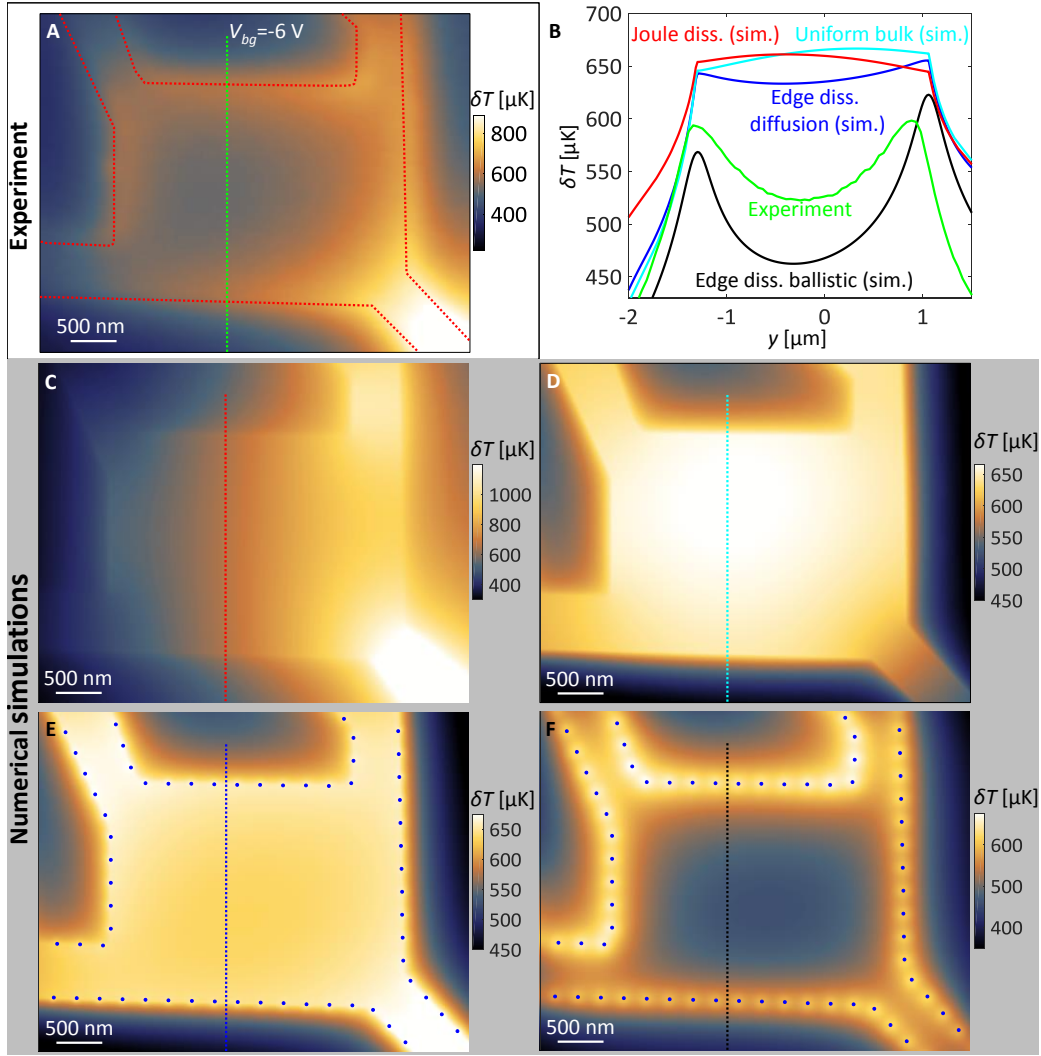

**Figure S6. Edges as dominant source of dissipation in hBN/graphene/hBN.**

(**A**) Thermal image $\delta T(x,y)$ of Sample 2 at tSOT flat-band condition as described in Figs. 4F,G but at $V_{bg}=-6$ V revealing dissipation along the graphene edges. Scan area 4.7 × 3.7 µm², pixel size 50 nm, scan-speed 200 ms/pixel, $h=20$ nm, $I_{dc}=1$ µA rms chopped at 35.5 Hz, $V_{tg}=-1.08$ V corresponding to $V_{tg}^{FB}$ at this $V_{bg}$. (**B**) Temperature profiles taken along the vertical dashed lines comparing Joule-heating simulation (**C** – red), uniform bulk heating simulation (**D** – cyan), edge point sources with diffusion model (**E** – blue), edge point sources with ballistic 2D phonons (**F** – black), and the experimental profile (**A** – green). (**C-E**) 3D heat flow simulations of surface temperature of graphene flake on SiO₂ substrate shaped as in experiment for three cases of heat sources. (**C**) Joule heating: A current of 1 µA is injected into the graphene through the right constrictions as in experiment. A sheet resistance of 180 Ω/□ is assigned to the graphene and the current density is solved. The current density is used as a heat source (resulting in a total heat load of 1.8 nW) to solve for the 3D heat flow problem. (**D**) Surface temperature from the solution of a heat flow problem with uniform power density dissipated in the graphene with the same total power as in (**C**). (**E**) Surface temperature from the solution of a diffusive heat flow problem with heat injected at discrete points (blue) along the edges with the same total power as in (**C**). (**F**) Surface temperature attained by superposition of thermal profiles of the form $T(r) \propto (1+(r/\ell)^2)^{-1/2}$ around each marked point (blue) representing the case of 2D ballistic phonons emitted from point sources along the sample edges.

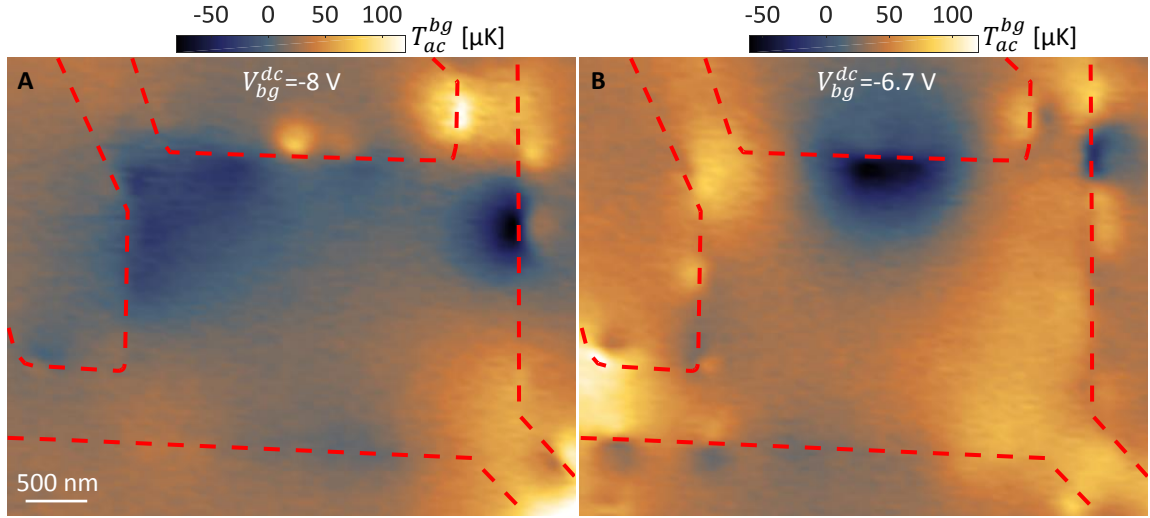

**Figure S7. Resolving resonant dissipation sources by *ac* back-gate voltage modulation $V_{bg}^{ac}$.**

**(A)** Thermal image $T_{ac}^{bg}(x,y)$ in presence of $I_{dc} = 6$ µA injected as indicated in Fig. 4F. An *ac* modulation $V_{bg}^{ac} = 0.1$ V rms at 322 Hz is applied to the back gate in addition to $V_{bg}^{dc} = -8$ V, $V_{tg} = -1.35$ V corresponding to $V_{tg}^{FB}$ at this $V_{bg}$. **(B)** Same as (**A**) at $V_{bg}^{dc} = -6.7$ V, $V_{tg} = -1.17$ V corresponding to $V_{tg}^{FB}$ at this $V_{bg}$. The bright and dark spots reveal the LSs that are shifted in and out of resonance by the back gate modulation $V_{bg}^{ac}$. Scan area $4.6 \times 3.7$ µm², pixel size 40 nm, scan-speed 60 ms/pixel, $h = 20$ nm.

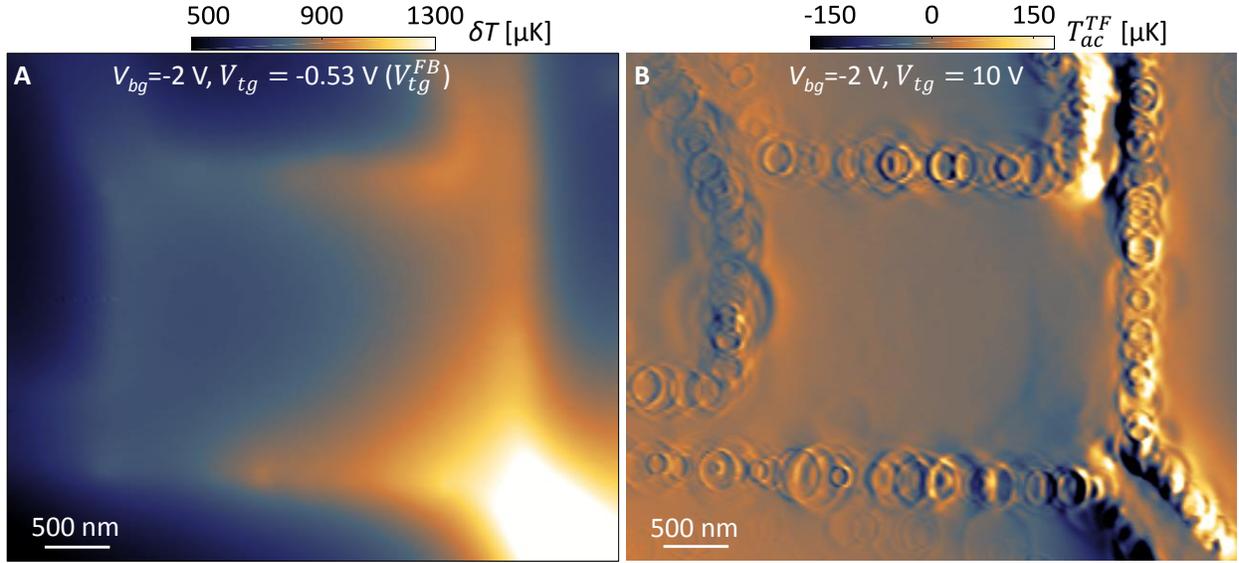

**Figure S8. Revealing dissipation from individual LSs by tip gating.**
**(A)** Thermal image $\delta T(x,y)$ at flat-band condition in presence of $I_{dc} = 1$ µA rms chopped at 35.5 Hz injected through the two right constrictions as indicated in Fig. 4F. Scan area 4.7 × 3.9 µm², pixel size 25 nm, scan-speed 200 ms/pixel, $h = 20$ nm, $V_{bg} = -2$ V, $V_{tg} = -0.53$ V corresponding to $V_{tg}^{FB}$ at this $V_{bg}$. **(B)** $T_{ac}^{TF}(x,y)$ thermal image away from flat-band condition revealing the individual resonant LSs at graphene edges in form of rings. The faint rings at the bottom of the image arise from the graphene edges in the adjacent chamber (see Fig. S4A). Scan area 4.7 × 3.9 µm², pixel size 20 nm, scan-speed 60 ms/pixel, $h = 20$ nm, $V_{bg} = -2$ V, $V_{tg} = 10$ V, $I_{dc} = 4$ µA rms chopped at 35.5 Hz.

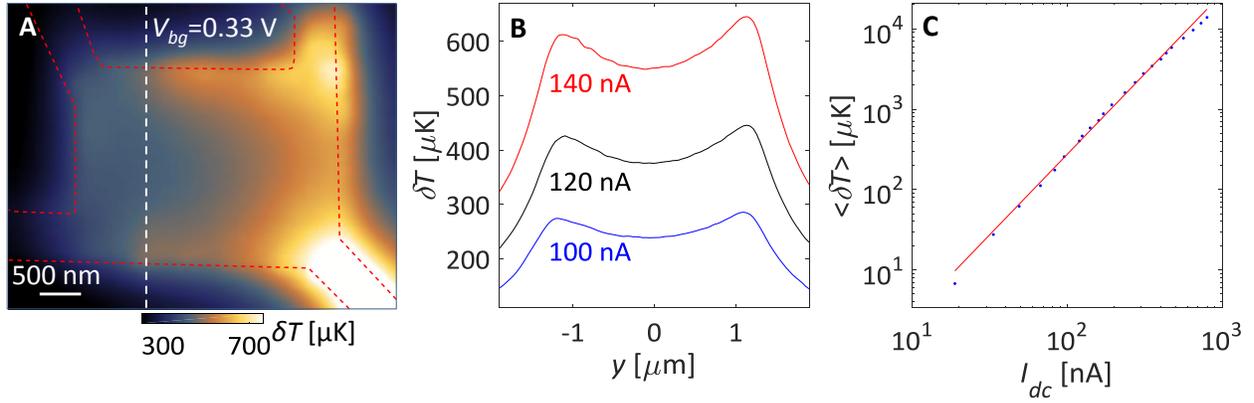

**Figure S9. Current dependence of the thermal map at flat-band condition.**

**(A)** Thermal image $\delta T(x,y)$ in presence of $I_{dc} = 120$ nA rms chopped at 35.5 Hz injected through the two right constrictions as indicated in Fig. 4F (the dotted line outlines the graphene edges). Scan area $4.7 \times 3.7$ µm², pixel size 60 nm, scan-speed 600 ms/pixel, $h = 20$ nm, $V_{bg} = 0.33$ V close to the CNP, $V_{tg} = -0.16$ V corresponding to $V_{tg}^{FB}$ at this $V_{bg}$. **(B)** $\delta T$ profiles along the white dashed line in **(A)** at several indicated values of $I_{dc}$. **(C)** Current dependence of the average temperature $\langle \delta T \rangle$ along the profiles (blue) and a quadratic fit $\langle \delta T \rangle \propto I_{dc}^2$ (red).

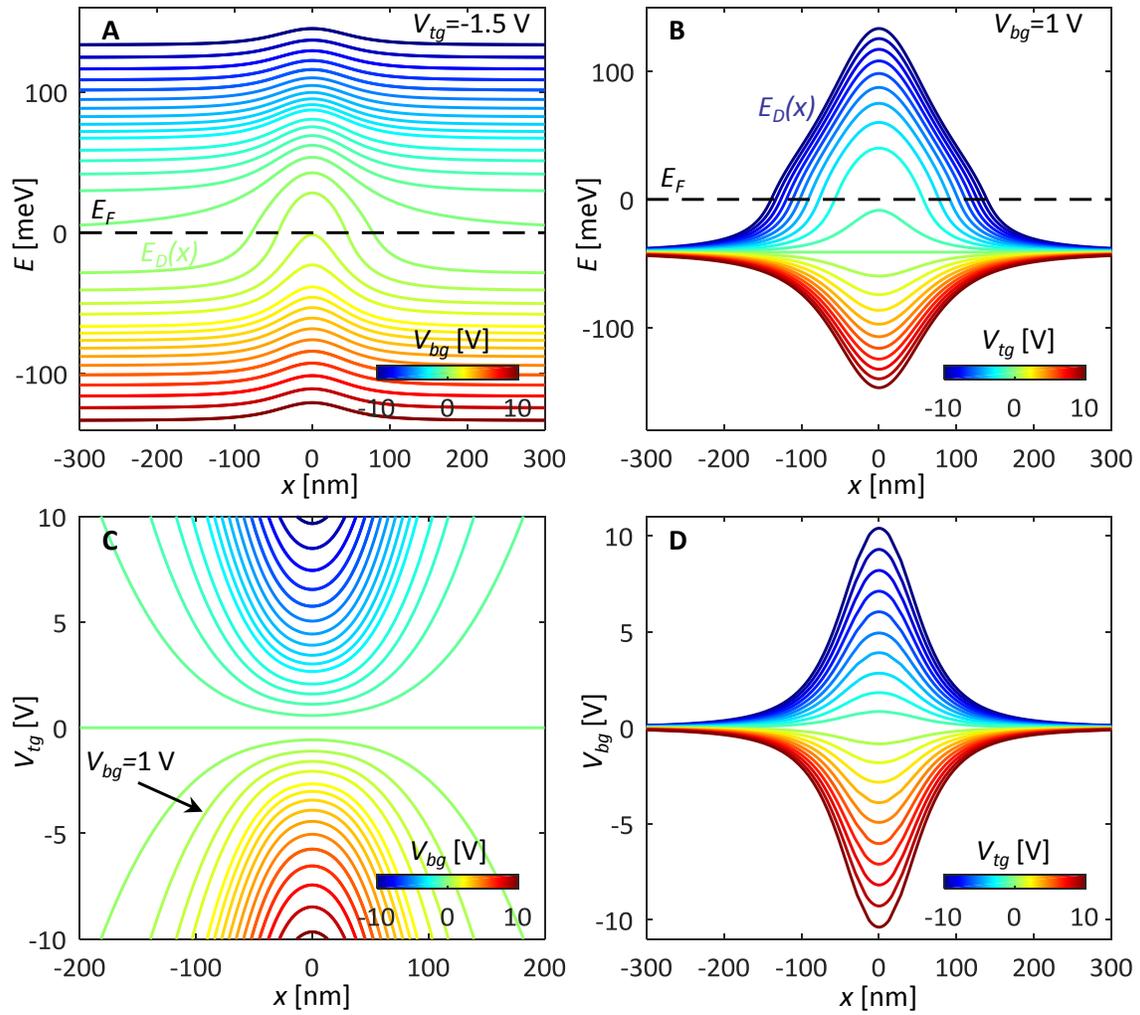

**Figure S10. Numerical simulations of tip-induced potentials and spectroscopy.**
(**A**) Band bending $E_D(x)$ induced in graphene by the tSOT located at $x = 0$ at potential $V_{tg} = -1.5$ V for various back gate voltages $V_{bg}$. (**B**) Band bending $E_D(x)$ in graphene at $V_{bg} = 1$ V and different values of $V_{tg}$. (**C**) Resonant dissipation traces describing the lateral displacement $x$ of the tSOT from the defect at which resonant conditions for inelastic scattering are met as a function of $V_{tg}$ for various values of $V_{bg}$ depicting the experimental results in Fig. S11B. (**D**) Resonant dissipation bell-shaped traces as a function of $V_{bg}$ for various values of $V_{tg}$ describing the experimental results in Fig. 4B and Fig. S11E.

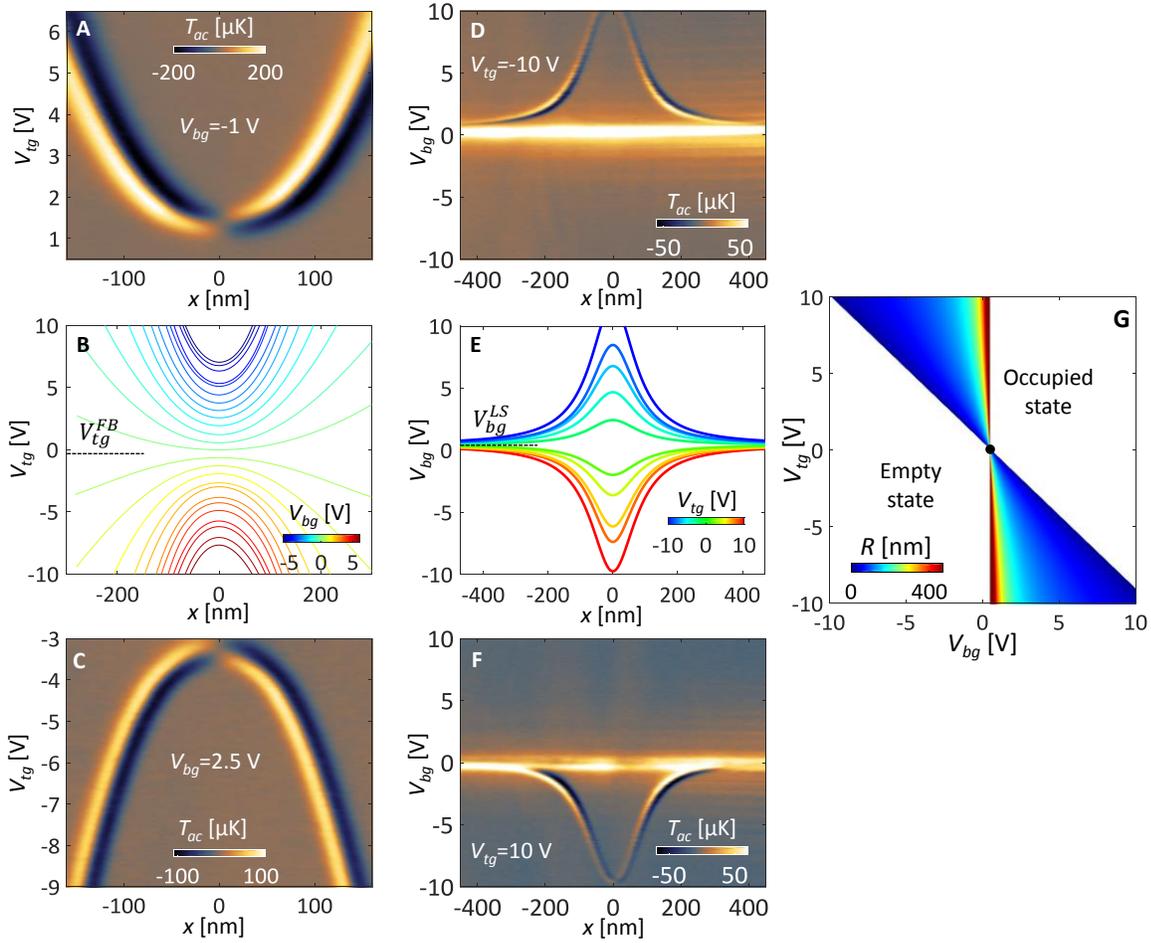

**Figure S11. Thermal spectroscopy of a single localized state.**
**(A)** Map of $T_{ac}(x)$ line scans through the center of defect 'C' upon varying $V_{tg}$ at constant $V_{bg} = -1$ V and $h = 20$ nm. Pixel width 3 nm, pixel height 50 mV, scan-speed 60 ms/pixel. **(B)** The resulting resonance traces $R(V_{tg})$ for various values of $V_{bg}$. The transition from convex to concave traces occurs at $V_{bg}^{LS}, V_{tg}^{FB}$. **(C)** Map of $T_{ac}(x)$ line scans as in (A) at $V_{bg} = 2.5$ V. **(D)** Map of $T_{ac}(x)$ line scans showing the bell-shaped resonance trace at $V_{tg} = -10$ V over the full range of $V_{bg} = \pm 10$ V at $h = 12$ nm, $x_{ac} = 2.7$ nm, and $I_{dc} = 6$ µA. The bright strip at $V_{bg} \approx 0$ V arises from enhanced heating at the nearby constriction, the resistance of which becomes very large close to CNP. **(E)** The resulting fitted bell-shaped resonance traces $R(V_{bg})$ for various values of $V_{tg}$ that change their polarity at $V_{bg}^{LS}, V_{tg}^{FB}$. **(F)** Same as (D) at $V_{tg} = 10$ V. The fact that only a single resonance trace is observed in (**D**) and (**F**) shows that the LS has only one energy level in the accessible range of $V_{tg}$ and $V_{bg}$. **(G)** The dissipation ring radius $R(V_{bg}, V_{tg})$ of the LS over the entire range of tip and back gate voltages with the $V_{bg}^{LS}, V_{tg}^{FB}$ point marked by black dot.

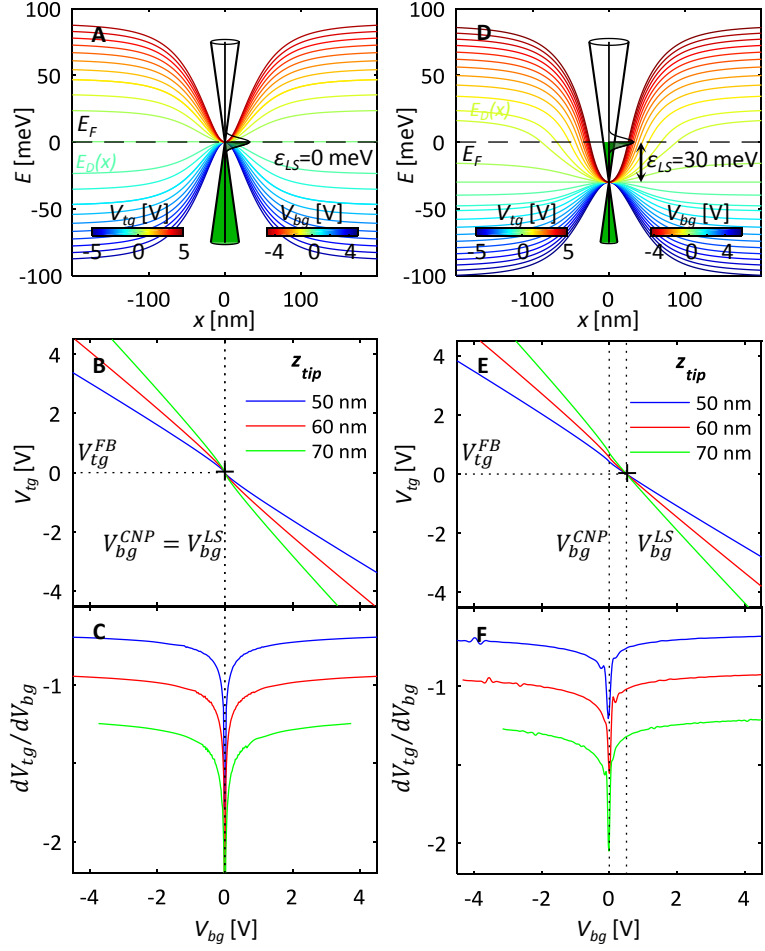

**Figure S12. Numerical simulation of resonance lines of LS.**
(**A**) Band bending $E_D(x)$ induced in graphene by the tSOT located at $x = 0$ at $z_{tip}$ = 60 nm above the graphene for critical $(V_{bg}, V_{tg})$ pairs such that $E_D(x = 0) = 0$ providing resonance conditions for inelastic scattering from LS with $\varepsilon_{LS} = 0$ (all other simulation parameters are identical to Fig. S10). Inset: Schematic Dirac cone with LS aligned with $E_F$. (**B**) The resulting $R = 0$ resonance curves for different $z_{tip}$ heights. The curves intersect at flat-band conditions determined by $V_{tg}^{FB}$ and $V_{bg}^{LS}$ (cross). (**C**) Derivative of the resonance curves showing the kink location determined by the local CNP, $V_{bg}^{CNP}$. For $\varepsilon_{LS} = 0$ the crossing point coincides with the kinks. (**D**) Same as (**A**) for the case of $\varepsilon_{LS} = 30$ meV. The resonance conditions are attained when $E_D(x = 0) = -30$ meV. (**E,F**) Same as (**B,C**) for the case of $\varepsilon_{LS}$ =30 meV. The kink and the crossing point are well separated and their difference, $V_{bg}^{LS} - V_{bg}^{CNP}$, provides a direct measure of $\varepsilon_{LS}$.

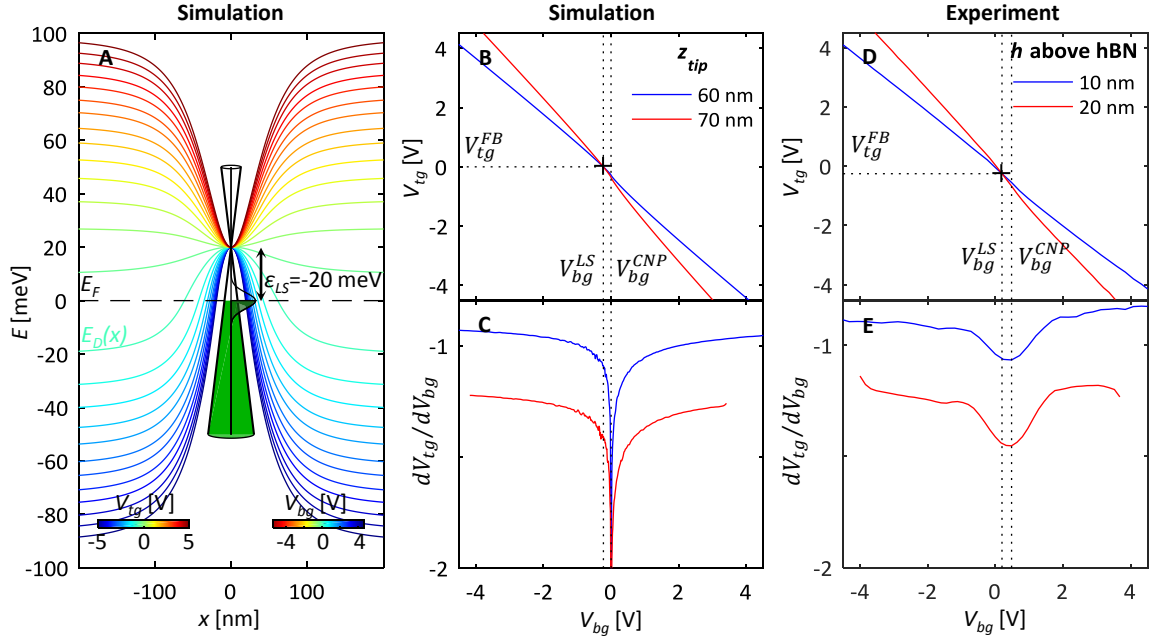

**Figure S13. Calculated and measured resonance lines of LS.**
**(A)** Calculated band bending $E_D(x)$ induced in graphene by the tSOT as in Fig. S12 for the case of $\varepsilon_{LS} = -20$ meV. Inset: Schematic Dirac cone with LS aligned with $E_F$ at resonance conditions. **(B,C)** The resulting calculated resonance curves for different $z_{tip}$ values **(B)** and the derivatives of the curves **(C)**, showing the intersection and the kink points $V_{bg}^{LS}$ and $V_{bg}^{CNP}$. **(D)** Experimental resonance lines of defect 'C' extracted from the data of Fig. 2D and a similar measurement performed at $h = 20$ nm above the hBN surface with otherwise the same measurement parameters, and their numerical derivatives **(E)**. A kink is apparent at the local change neutrality point, $V_{bg}^{CNP} = 0.48$ V, defined by the minima in **(E)**. Experimental curves intersect at $V_{bg}^{LS} = 0.2$ V and $V_{tg}^{FB} = -0.26$ V (indicated by the cross in **(D)**), resulting in $\varepsilon_{LS} = -22$ meV.

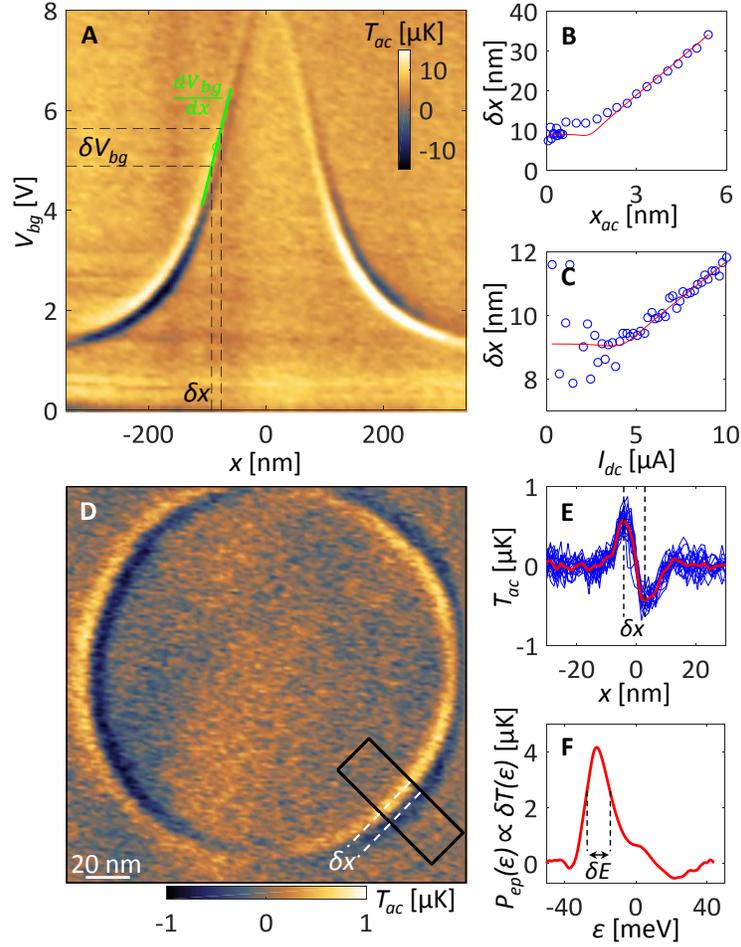

**Figure S14. Estimating the spectral width of the localized state.**
**(A)** Bell-shaped $T_{ac}$ dissipation trace of defect 'B' at $V_{tg} = -10$ V, $h = 20$ nm, $x_{ac} = 2.7$ nm, and $I_{dc} = 3$ µA. The slope of the trace $dV_{bg}/dx$ (at $V_{bg} = 5.3$ V, green line) translates the width of the ring, $\delta x$, into corresponding width $\delta V_{bg}$ and $\delta E$ of the LS. **(B)** Width $\delta x$ of the dissipation ring vs. the tip vibration rms amplitude $x_{ac}$ at $I_{dc} = 3$ µA (red curve is a guide to the eye). **(C)** Width $\delta x$ of the dissipation ring vs. $I_{dc}$ at $x_{ac} = 0.5$ nm. **(D)** $T_{ac}$ dissipation ring at $V_{tg} = -10$ V, $V_{bg} = 5.3$ V, $h = 20$ nm, $x_{ac} = 0.5$ nm, and $I_{dc} = 2$ µA (Scan area $180 \times 180$ nm², pixel size 1.2 nm, scan-speed 600 ms/pixel). **(E)** Several line-cuts in the rectangle in (C) (blue, x-axis parallel to long edge of the rectangle), and the average of the profiles (red) determining the width of the ring $\delta x = 7$ nm. **(F)** The corresponding $\delta T(\varepsilon) \propto P_{ep}(\varepsilon)$, with $\delta E = 13$ meV.

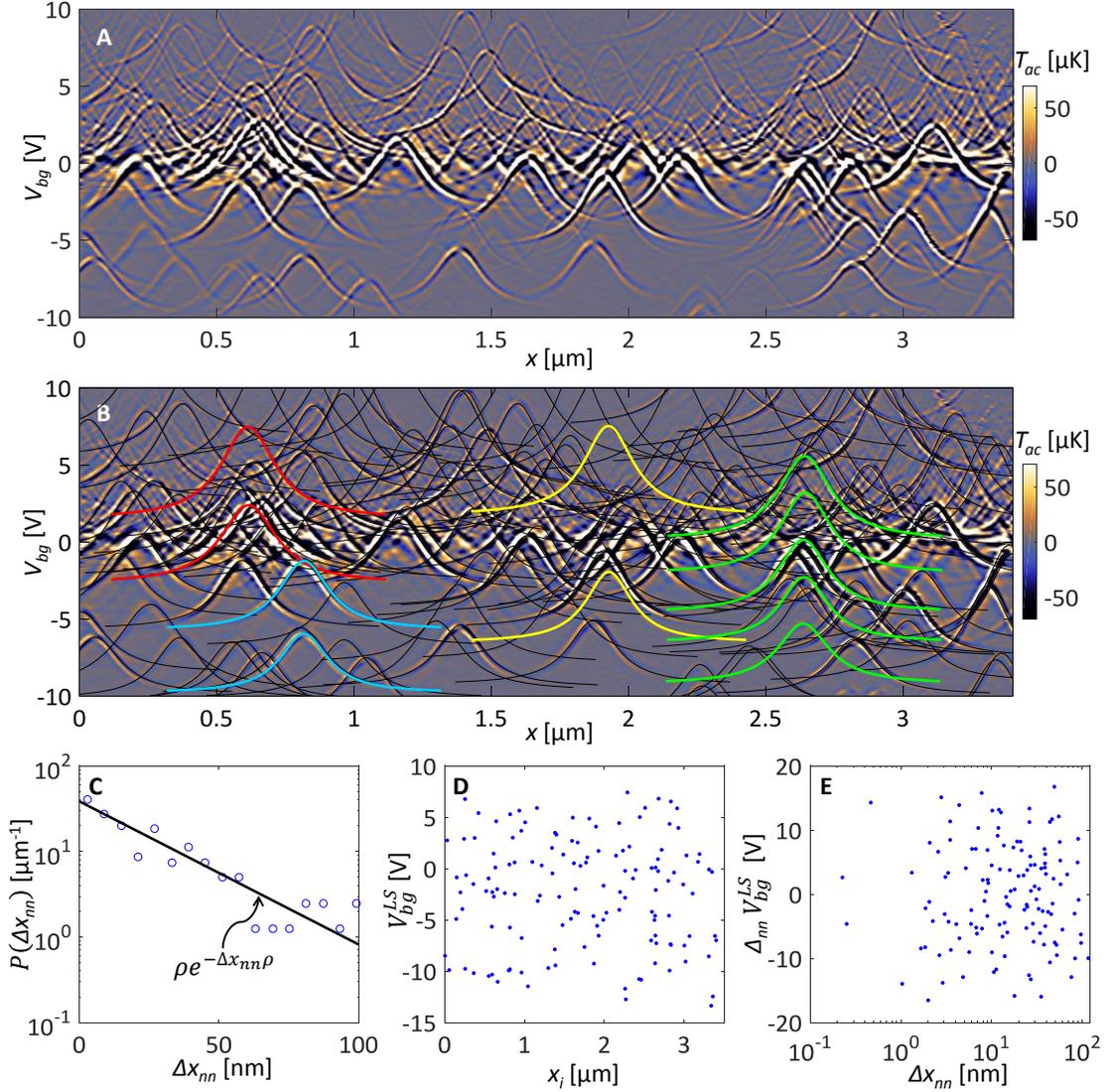

**Figure S15. Spectroscopic thermal imaging of the localized edge states.**
**(A)** Map of $T_{ac}(x)$ line scans along the bottom edge of the graphene sample (reproduced from Fig. 4D) vs. $V_{bg}$ at $V_{tg} = -10$ V showing bell-shaped resonant dissipation traces of the individual LSs. **(B)** Numerical fits (black) to the bell-shaped traces of 135 clearly resolved LSs. The asymptotic values of the traces $V_{bg}^{LS}$ describe the energy $E_{LS}$ of the LSs. The colored traces (red, blue and yellow) show examples of pairs of adjacent neighbors with significantly different energy $V_{bg}^{LS}$. The green traces show an example of five LSs within 8 nm interval. **(C)** Probability density of the nearest neighbor distances $\Delta x_{nn}$ between the LSs (20 bins) and the theoretical exponential decay for random distribution (black). **(D)** $V_{bg}^{LS}$ values of the resolved states vs. their location $x_i$ along the edge showing large uncorrelated variation in the energy of the LSs. **(E)** The difference in $V_{bg}^{LS}$ between the nearest neighbors $\Delta_{nn} V_{bg}^{LS}$ vs. the distance $\Delta x_{nn}$ between the neighboring LSs.

**Movie Captions**

**Movie S1. Thermal imaging of dissipation in hBN encapsulated graphene**

Scanning nanothermometry movie of $T_{ac}(x,y)$ of 4×4 µm² hBN/graphene/hBN sample. The tSOT of 33 nm effective diameter is scanned at $h = 20$ nm above the sample surface in presence of a *dc* current $I_{dc} = 3$ µA applied between the top and the bottom-right constrictions at back gate voltage $V_{bg} = -2$ V. The tSOT is attached to a tuning fork and oscillates at 36.9 kHz with rms amplitude $x_{ac} = 2.7$ nm parallel to the surface at an angle of 12° relative to the $x$ axis, providing sensitive $T_{ac}(x,y)$ imaging. The movie shows a sequence of images as the tip potential $V_{tg}$ is incremented from 0 to 9 V. Numerous dissipation rings appear along the graphene edges and expand as $V_{tg}$ increases. In addition, three isolated dissipation rings form and expand inside the sample revealing rare atomic defects in the bulk of graphene. The bulk and edge defects form sharp localized states acting as point source emitters of phonons due to resonant inelastic scattering of electrons. Scan area 5.5 × 5 µm², pixel size 18.3 nm, scan-speed 20 ms/pixel.

**Movie S2. Thermal imaging of resonant dissipation of three bulk localized states**

Scanning nanothermometry movie of $T_{ac}(x,y)$ of the central part of hBN/graphene/hBN sample. Three resonant dissipation rings appear and expand upon increasing $V_{tg}$ from 2 to 10 V revealing inelastic electron scattering from three individual atomic defects. The small differences in the onset potential of the rings can be ascribed to long-range substrate potential disorder modifying the local CNP. $V_{bg} = -1$ V, $I_{dc} = 3$ µA, $h = 20$ nm, $x_{ac} = 2.7$ nm, scan area 3.5 × 2.8 µm², pixel size 17.5 nm, scan-speed 20 ms/pixel. The slightly non-monotonic expansion of the rings is a result of fluctuations in the tSOT scanning height.

**Movie S3. Principle of spectroscopic thermal imaging of localized states**

Animation describing thermal spectroscopy of the dissipation process due to resonant inelastic scattering of electrons from a localized state. A single LS with an energy level at $E_{LS}$ pinned to the Dirac energy $E_D$ is located at $x = 0$ marked by a sharp peak in $P_{ep}(\varepsilon)$. The Fermi energy $E_F$ (dashed) is determined by the back gate voltage $V_{bg}$. Electrons are injected from left at energy $E_e$ (red arrows). The potential $V_{tg}$ of the moving tSOT induces local band bending of the graphene Dirac energy $E_D(x)$ (blue) which shifts the energy of the LS. Inelastic scattering of the electrons occurs when the tip-induced potential brings the LS into resonance, $E_{LS} = E_e$, resulting in phonon emission from the defect (red arcs) which is detected by the tSOT as a sharp $\delta T(x)$ thermal signal (red curve) when positioned at $x = \pm R$. In 2D thermal imaging this results in a ring of enhanced $\delta T(x,y)$ with its center at the LS. The $\delta T_{on}(x)$ curve (dashed) describes the local temperature that would be measured by the tSOT if the LS stayed at resonance constantly. The band bending curve $E_D(x)$ (blue) was calculated using the electrostatic simulations as described in Section 6.

**Movie S4. Spectroscopic dissipation traces of a single bulk localized state vs. $V_{tg}$**

Spectroscopic nanothermometry movie of $T_{ac}(x)$ of defect 'C'. The tSOT is scanned along a line through the center of the LS upon varying $V_{tg}$. The different frames show the evolution of the resonant dissipation trace with $V_{bg}$. The curvature of the trace flips at $V_{bg}^{LS}$. $I_{dc} = 3$ µA, $h = 20$ nm, pixel width 3 nm, pixel height 50 mV, scan-speed 60 ms/pixel.

**Movie S5. Spectroscopic bell-shaped dissipation traces of a single bulk localized state vs. $V_{bg}$**

Spectroscopic nanothermometry movie of $T_{ac}(x)$ of defect 'C'. The tSOT is scanned along a line through the center of the LS upon varying $V_{bg}$. The different frames show the evolution of the resonant trace upon changing the tip potential $V_{tg}$. The curvature of the trace flips at $V_{tg}^{FB}$. The bright $V_{tg}$-independent signal in a form of a strip around $V_{bg} \approx 0$ V is the result of heating of the nearby constriction that peaks when $V_{bg}$ reaches CNP of the constriction. $I_{dc} = 3$ µA, $h = 20$ nm, pixel width 5 nm, pixel height 30 mV, scan-speed 20 ms/pixel.

**Movie S6. Spectroscopic bell-shaped dissipation traces of edge localized states vs. $V_{bg}$**

Spectroscopic nanothermometry movie of $T_{ac}(x)$ of edge LSs. The tSOT is scanned along the bottom edge of the graphene upon varying $V_{bg}$, revealing numerous resonant bell-shaped dissipation traces. The different frames show the evolution of the resonant traces with $V_{tg}$. The resonant dissipative edge states are observed at all values of $V_{bg}$ with very large variations in $V_{bg}^{LS}$. $I_{dc} = 3$ µA, $h = 20$ nm, pixel width 4 nm, pixel height 100 mV, scan-speed 60 ms/pixel. High-pass filtering was applied to emphasize the pertinent bell-shaped resonance traces.

**Movie S7. Spectroscopic dissipation traces of edge localized states vs. $V_{tg}$**

Spectroscopic nanothermometry movie of $T_{ac}(x)$ of edge LSs. The tSOT is scanned along the bottom edge of the graphene upon varying $V_{tg}$, revealing numerous resonant dissipation traces. The different frames show the evolution of the resonant traces with $V_{bg}$. The resonant edge states are observed at all values of $V_{bg}$ with very large variations in $V_{bg}^{LS}$ at which the curvature of trace originating from a given LS flips. $I_{dc} = 3$ µA, $h = 20$ nm, pixel width 10 nm, pixel height 100 mV, scan-speed 20 ms/pixel.